\documentclass[final,3p,times]{elsarticle}

\usepackage{graphicx} 
\usepackage{caption}
\usepackage{subcaption}
\usepackage{amsmath, amsthm}
\usepackage{amsfonts}
\usepackage{import}
\usepackage{bbm}
\usepackage{algorithm}
\usepackage{algpseudocode}
\usepackage{comment}
\usepackage{siunitx}
\usepackage{pdfpages}
\usepackage{mathtools}
\usepackage{multicol}
\usepackage{bm}
\usepackage{hyperref}
\usepackage{tikz}
\usetikzlibrary{shapes.geometric}
\usepackage{xcolor}
\usepackage{colortbl}
\usepackage{booktabs}
\usepackage[most]{tcolorbox}
\usepackage{pdfcomment}
\usepackage{dirtytalk}
\usepackage{algpseudocode}
\usepackage[utf8]{inputenc}
\usepackage{cleveref}
\usepackage{makecell}
\usepackage{pifont}
\usepackage[disable]{todonotes}  
\usepackage{soul}

\usepackage{tcolorbox}
\usepackage{soul}
\usepackage{xcolor}

\newif\ifhighlight
\highlighttrue   
\highlightfalse
\ifhighlight
  
\else
  
\fi

\ifhighlight
  \newtcolorbox{hlblock}{
    colback=yellow,
    colframe=yellow,
    boxrule=0pt,
    arc=2pt,
    left=4pt, right=4pt, top=4pt, bottom=4pt
  }
\else
  \newenvironment{hlblock}{}{}
\fi

\newcommand{\pyhex}[2][0.7ex]{
  \tikz[baseline=-0.75ex]{%
    \fill[#2] 
      (0:#1) -- (60:#1) -- (120:#1) -- (180:#1) -- (240:#1) -- (300:#1) -- cycle;
  }%
}

\newcommand{\BO}{\mathcal{O}}

\newcommand{\EX}{\mathbb{E}}

\newcommand{\dt}{\Delta t} 

\newcommand{\intv}{\int_{-\infty}^\infty}

\definecolor{KULeuvenLichtblauw}{RGB}{0, 0, 255}
\usetikzlibrary{positioning} 

\begin{document}

\begin{frontmatter}


\title{
A kinetic-diffusion Monte Carlo-based particle-level fluid-kinetic decomposition for neutral transport simulations
}
\author{Zhirui Tang\corref{cor1}\fnref{lab1}} 
\author{Niels Horsten\fnref{lab2,lab3}}
\author{Giovanni Samaey\corref{cor1}\fnref{lab1}}

\cortext[cor1]{Corresponding author: zhirui.tang@kuleuven.be; giovanni.samaey@kuleuven.be}

\affiliation[lab1]{organization={Department of Computer Science, KU Leuven},
            city={Leuven},
            postcode={3001}, 
            country={Belgium}}
            
\affiliation[lab2]{organization={Department of Mechanical Engineering, KU Leuven},
            city={Leuven},
            postcode={3000}, 
            country={Belgium}}

\affiliation[lab3]{organization = {Institute of Mechanical, Materials and Civil Engineering, UCLouvain},
city = {Louvain-la-Neuve},
postcode = {1348},
country = {Belgium}}

\begin{abstract}
Neutrals in the plasma edge are commonly modeled by kinetic equations, with quantities of interest given by macroscopic quantities such as density, velocity, and temperature. 
In reactor-relevant regimes, fully kinetic descriptions solved by Monte Carlo (MC) methods, while accurate, become computationally expensive, whereas fluid-limit approximations are computationally more efficient but may lose accuracy due to boundary effects or low-collisional regimes. 
Hybrid fluid–kinetic approaches aim to combine the strengths of both descriptions. However, existing simulation methods face challenges, including interface handling in domain decomposition, unphysical assumptions, and iterative coupling in distribution decomposition.
In this work, we propose a distribution-decomposition hybrid model constructed at the particle level based on the kinetic–diffusion Monte Carlo (KDMC) method. 
The model inherits key properties of KDMC: it is asymptotic-preserving and does not require iterative coupling between the fluid and kinetic components.
To improve the accuracy of the fluid-part quantities estimation, a Navier–Stokes–type fluid system is derived via Hilbert-Chapman–Enskog expansions, tailored for KDMC. 
In the considered one-dimensional tests, the resulting fluid system has comparable accuracy to the AFN model used in SOLPS-ITER while requiring substantially fewer nonlinear iterations.
Additionally, a tunable reflective boundary condition is introduced that allows balancing accuracy and efficiency. 
The model exhibits at least 500 times speedup over the kinetic MC, while maintaining relative $L_2$ errors around $10\%$ in a charge-exchange (CX) dominant test case. 
In non-CX-dominant regimes, the accuracy becomes increasingly sensitive to boundary treatment due to the inherent limitations of the fluid approximation near the boundary, motivating further refinement of the KDMC boundary conditions.

\end{abstract}


\end{frontmatter}



\section{Introduction}

In the plasma edge of nuclear fusion reactors, neutral transport is often modeled by kinetic equations. The neutral-related quantities of interest (QoIs) are typically macroscopic quantities, such as density, velocity, and temperature, relating to the first three moments of the solution of the kinetic equation. 
In reactor-relevant regimes, Monte Carlo (MC) particle tracing schemes and MC estimators are commonly employed to solve the kinetic equations and estimate the QoIs, respectively~\cite{reiterEIRENEB2EIRENECodes2005a, mortierStudySourceTerm2024}.
While accurate, these methods become computationally expensive in high-collisional regimes, whereas fluid-limit are fast but may lose accuracy in low-collisional regimes or due to boundary effects.

Hybrid fluid-kinetic models that combine kinetic and fluid descriptions, therefore, offer a promising alternative.
These approaches were initially developed in the context of radiation transport~\cite{fleckRandomWalkProcedure1984b}, and were subsequently applied to neutron transport~\cite{BORGERS1992285}, rarefied gas dynamics~\cite{sunHybridContinuumParticle2004}, among other applications.
In neutral transport of the plasma-edge simulations, hybrid models are designed to balance the accuracy of kinetic models in low-collisional regimes and in the presence of boundary effects with the efficiency of fluid limits in high-collisional regimes where they are accurate. 

Hybrid fluid-kinetic approaches can be roughly classified into two types. 
The first type is domain decomposition, where the simulation domain is divided into kinetic and fluid regions, and different methods are applied to each part, see, for instance, the consecutive works~\cite{blommaertSpatiallyHybridFluidkinetic2019a, horstenApplicationSpatiallyHybrid2021a, vanuytven2022}.  
A major difficulty of this approach is that an artificial boundary condition (BC) must be imposed inside the original computational domain, and the interfaces separating the kinetic and fluid regions may have complex geometries. In time-dependent simulations or iterative procedures, these interfaces can move as a function of time. Handling such moving boundaries is a challenge~\cite{karneyModelingNeutralPlasma1998}.
Moreover, the determination of the interface between the fluid and kinetic regions can be challenging and has a significant impact on the accuracy of the method~\cite{degondFluidSimulationsLocalized2012a, degondSmoothTransitionModel2005a, letallecCouplingBoltzmannNavier1997}.
More domain decomposition schemes in other applications can be found in~\cite{degondMultiscaleKineticFluid2010, crouseillesHybridKineticFluid2004a, tiwariParticleParticleHybrid2009a, kolobovUnifiedSolverRarefied2007}.

The second type is distribution decomposition, where the neutral distribution, and hence the original kinetic equation, is decomposed into a kinetic and a fluid part. Both parts are solved in the entire domain.
Generally, the kinetic part model is solved by kinetic MC, and the fluid part is solved by taking the fluid limit.
The first hybrid model for the neutrals in the plasma edge belongs to this type, and was theoretically elaborated in 1998~\cite{karneyModelingNeutralPlasma1998}. It was subsequently implemented in the SOLEDGE2D-EIRENE code suite~\cite {valentinuzziTwophasesHybridModel2019b}.
In this scheme, two user-defined source and sink terms are added to the kinetic and fluid part models, such that a higher weight is assigned to the fluid part model in high-collision regimes, and to the kinetic part model in low-collisional regimes. 
However, this approach faces the difficulties of selecting the artificial source and sink terms.
Later, another hybrid model~\cite{horstenHybridFluidkineticModel2020a} of the distribution decomposition type was proposed and implemented in the SOLPS-ITER code suite~\cite{horstenHybridFluidKinetic2020}. It relies on the assumption that the entire mass and momentum of particles are concentrated in the fluid part. As such, the kinetic correction term carries zero mass and momentum. 
This unphysical approach complicates the sampling, since particles with negative mass are now needed. 
Computing a small correction term from a population of many positive and negative particles whose masses nearly cancel out leads to cancellation errors.
Additionally, in both models discussed above, the fluid and kinetic parts are coupled, such that they have to be solved iteratively. 
Further examples of function decomposition schemes in different applications can be found in~\cite{dimarcoHybridMultiscaleMethods2008b, katsoulakisMultiscaleCouplingsPrototype2004, letallecCouplingBoltzmannNavier1997a}.

To overcome the difficulties mentioned above, we propose a distribution-decomposition hybrid model constructed at the particle-simulation level, making use of the kinetic–diffusion Monte Carlo (KDMC) method~\cite{mortierKineticDiffusionAsymptoticPreservingMonte2022}.
KDMC is an asymptotic-preserving \cite{jinEfficientAsymptoticPreservingAP1999c, dimarcoHybridMultiscaleMethods2008c, crestettoParticleMicromacroDecomposition2018a, lovbakAcceleratedSimulationBoltzmannBGK2023} hybrid particle simulation scheme that generates particle trajectories by combining kinetic steps from kinetic MC with diffusive steps that represent the cumulative effect of many individual kinetic steps. 
It behaves as kinetic MC in low-collisional regimes and converges to the diffusive fluid limit in highly collisional regimes. 
In intermediate regimes, KDMC results in a bias that can be removed with a multilevel extension, as developed in~\cite{mortier2022a}.
The resulting hybrid model is asymptotic-preserving, since its basis, KDMC, is. The kinetic and fluid parts are not coupled due to the natural splitting at the particle level. Consequently, no iterative scheme is required, in contrast to the aforementioned distribution-decomposition hybrid models.
\todo{*}

Being a fully particle-based hybrid method, KDMC avoids many of the problems of alternative hybrid methods. Still, KDMC has a few drawbacks, which will be circumvented by the method proposed in this paper.
One problem of KDMC is that the trajectories do not give information about individual collisions during diffusive steps, rendering classical MC estimation techniques inappropriate.
In \cite{mortierEstimationPostprocessingStep2022}, it was proposed to estimate the contributions of the diffusive steps to the QoIs by performing a time step of a time-dependent fluid equation, in which the initial condition and the simulation time are estimated from diffusive steps in the KDMC simulation. However, estimating the initial condition involves point estimation,
which typically exhibits a large variance~\cite{luxMonteCarloParticle2018a}.  
In addition, estimating the simulation time introduces an extra error with variance and bias in a heterogeneous background.
A full error analysis of KDMC with the estimation technique can be found in \cite{tangAnalysisKineticdiffusionMonte2025}.
As the main contribution in the proposed hybrid model, we estimate the contribution of the diffusive steps to the QoIs by solving a steady-state fluid model, avoiding the need to estimate the initial condition and the simulation time.
Instead, only an unknown source term in the fluid part of our hybrid model must be estimated. 
We prove that this source term can be estimated along the kinetic steps, which can reuse the QoIs estimation of the kinetic part and avoid the point estimation. Therefore, a lower variance is achieved with less cost.
As a result, our method provides both reduced bias and variance.
\todo{Removed the paragraph about how to derive the model. 
Added the advantages of hybrid model in the * position}

Besides these contributions, we make three additional extensions.
First, we incorporate the ionization process in KDMC, which so far has only included charge exchange collisions.
To this end, we derive a fluid equation governing the diffusive steps in KDMC following~\cite{maesHilbertExpansionBased2023a}, but with the ionization term included.
Second, the resulting fluid equation is a single density equation. 
However, it has been shown in~\cite{horstenComparisonFluidNeutral2016a} that such a fluid limit is not sufficiently accurate for plasma edge simulations. 
Therefore, following~\cite{horstenComparisonFluidNeutral2016a} but tailored for KDMC, we extend the fluid model to a Navier–Stokes-type fluid system based on the Hilbert-Chapman–Enskog expansion, including the density, momentum, and energy equations, which provides a much more accurate description.
Finally, we implemented outflow and reflective BCs.
We propose a parametrized scheme for the reflective BC that interpolates between the kinetic and fluid BCs, and that can be used to balance accuracy with computational speed.

In this work, we restrict our attention to the one-dimensional setting, with 
$x \in \mathbb{R}$ and $v \in \mathbb{R}$. 
While the present work is restricted to one spatial dimension, the KDMC-based particle decomposition itself is not intrinsically one-dimensional. Extension to higher spatial dimensions requires the derivation and assessment of suitable multidimensional fluid closures and boundary treatments, and is left for future work.

The rest of this paper is organized as follows. The kinetic equations and the quantities of interest are introduced in Sec.~\ref{sec: eq and QoI}. 
Since the proposed fluid system forms the basis of both KDMC and the hybrid model, 
it is first presented in Sec.~\ref{sec: fluid system}. 
The KDMC method, including the treatment of the ionization collision, is then described in Sec.~\ref{sec: KDMC}. 
The proposed hybrid model is constructed in Sec.~\ref{sec: hybrid model}. 
The BCs of the fluid model and hybrid model are discussed in Sec.~\ref{sec: bc}.
Then, in Sec.~\ref{sec: num}, numerical experiments are conducted to compare the proposed fluid model, the fluid model in \cite{horstenComparisonFluidNeutral2016a}, the hybrid model, and the scheme in \cite{mortierEstimationPostprocessingStep2022}.
Finally, conclusions are drawn in Sec.~\ref{sec: conclusion}.


\section{Kinetic equations and quantities of interest}\label{sec: eq and QoI}
In this section, we first introduce the discussed steady-state linear kinetic equation 
and the corresponding time-dependent equation used in the MC particle simulation. 
The connection between the former and the latter equation will be exploited in the derivation of the proposed hybrid model in Sec.~\ref{sec: hybrid model}.

\subsection{Steady-state kinetic equations}\label{sec: kinetic eq}
Let $\phi(x,v)$ be the steady-state neutral distributions at position $x$ with velocity $v$.
The steady-state kinetic equation, which models the evolution of neutral particles colliding with the plasma (ions and electrons), is then
\begin{equation}\label{eqn: ss-eq}
    v\partial_x \phi(x,v) = R_{cx}(x)\left(M_p\left( v\, | \, x\right)\intv \phi(x,v')dv'-\phi(x,v)\right) - R_i(x) \phi(x,v)+s(x,v),
\end{equation}
where the source term $s(x,v)=R_r(x) M_p\left(x,v\right) n_p(x)$. In this model, the charge-exchange, electron impact ionization, and radiative recombination collisions are considered. The corresponding collision rates are $R_{cx}(x), R_i(x)$, and $R_r(x)$, respectively. The plasma distribution $M_p\left(x,v\right)$ is typically assumed to be a perfect drifting Maxwellian for neutral simulations \cite{reiterEIRENEB2EIRENECodes2005a}, and its normalization reads
\begin{equation}\label{eqn: maxwellian dist.}
    M_p\left( v\, | \, x\right)=M\left(v \mid u_p(x), T_p(x)\right) = \frac{\sqrt{m}}{\sqrt{2\pi T_p(x)}}\exp\left(\frac{|v-u_p(x)|^2}{2T_p(x)/m}\right),
\end{equation}
where the given functions $u_p(x)$ and $T_p(x)$ are the macroscopic velocity and temperature of ions, respectively, assuming equal ion and electron temperatures. The constant $m$ is the mass of the ion and also the neutral particles, since we neglect the mass of the electron. 

\subsection{Time-dependent kinetic equation}\label{sec: t kinetic eq}
The neutral particle is modeled by the steady-state equation \eqref{eqn: ss-eq}. 
In computer simulations, however, a time-dependent equation is used if a particle-based method (such as the kinetic MC or KDMC) is applied, 
since there is no steady-state behavior for a single particle, and the system evolves toward the steady-state from an initial state. 
As in~\cite{mortierStudySourceTerm2024,lapeyreIntroductionMonteCarloMethods2003b}, we define the time-dependent neutral distribution $f(x,v,t)$ at time $t$ by $\int_0^\infty f(x,v,t) dt = \phi(x,v)$
and solve 
\begin{equation}\label{eqn: t kinetic}
    \partial_t f(x,v,t) + v\partial_x f(x,v,t) =  R_{cx}(x)\left(M_p(v \mid x) \intv f(x,v',t) dv'  - f(x,v,t) \right) -R_i(x) f(x,v,t), 
\end{equation}
with the initial condition $f(x,v,t=0)=s(x,v)$ and the property
\begin{equation}\label{eqn: steady-state phi initial condition}
    f(x,v,t=\infty)=0,
\end{equation} 
due to the sink term $-R_i(x) f(x,v,t)$.
Integrating \eqref{eqn: t kinetic} over time $t$ from $0$ to $\infty$, we have
\begin{equation}\label{eqn: temp1}
    f(x,v,t=\infty) - f(x,v,t=0) + v\partial_x \phi(x,v) = R_{cx}\left(M_p(v \mid x) \intv \phi(x,v) dv  - \phi(x,v)\right) -R_i \phi(x,v).
\end{equation}
Substituting the initial condition $f(x,v,t=0)=s(x,v)$ and the property \eqref{eqn: steady-state phi initial condition} into Eq.~\eqref{eqn: temp1}, we obtain exactly the steady-state equation \eqref{eqn: ss-eq} we want to solve. 
With the time-dependent equation \eqref{eqn: t kinetic}, 
we sample the initial state of the particle system from the source term $s(x,v)$ and simulate all particles until they die (ionize). 
We refer to 
\cite{reiterEIRENEB2EIRENECodes2005a, tangAnalysisKineticdiffusionMonte2025} for the time-dependent MC simulation.

\subsection{Quantities of interest}\label{sec: QoIs}
In plasma edge simulations, instead of the neutral distribution $\phi(x,v)$, one is typically interested in the steady-state macroscopic quantities of neutrals: density $n(x)$, velocity $u(x)$, and temperature $T(x)$, defined as
\begin{equation}\label{eqn: QoIs}
    n(x)=m_0(x), \quad u(x)=\frac{m_1(x)}{n(x)}, \quad \text{and} \quad T(x)=m\left(\frac{m_2(x)}{n(x)}-u^2(x)\right)
\end{equation}
in which the three time-integrated moments $m_l$ with $l=0,1,2$ are defined as
\begin{align}\label{eqn: moments}
     m_l(x) &= \int_0^{\infty} \intv v^lf(x,v,t)dv dt
\end{align}
where the time integral from $0$ to $\infty$ is consistent with the transformation $\int_0^\infty f(x,v,t) dt = \phi(x,v)$ in Sec. \ref{sec: t kinetic eq}.

\section{Fluid system based on diffusive scaling assumption}\label{sec: fluid system}
Multiple fluid neutral models have been developed over the last few decades, in which the advanced fluid neutral (AFN) model~\cite{horstenDevelopmentAssessment2D2017, horstenComparisonFluidNeutral2016a} has been implemented in SOLPS-ITER and applied successfully in plasma edge simulations. 
AFN is a Navier-Stokes-type model derived from Eq.~\eqref{eqn: ss-eq} based on a hydrodynamic assumption, 
including neutral density, momentum, and energy equations, using the Chapman-Enskog expansion. 
It is shown that adding the energy equation is necessary to further reduce the model error (between the kinetic equation \eqref{eqn: ss-eq} and the derived fluid system). 

In this section, we derive an alternative Navier-Stokes-type fluid system, as done in the AFN model, but tailored for KDMC. This system is based on the diffusive scaling assumption introduced below and serves as a closure consistent with the diffusive steps of KDMC. It will form the basis for the extensions developed in the proposed hybrid method.

We first derive the density equation by means of a Hilbert expansion, following \cite{mortierKineticDiffusionAsymptoticPreservingMonte2022, maesHilbertExpansionBased2023a}, with the inclusion of the ionization term.
Afterwards, we derive the momentum and the energy equations, making use of the Chapman-Enskog framework as the AFN, but starting from the neutral distribution obtained from the Hilbert expansion.
\todo{To clear the innovation from AFN, I explicitly say here that we do use the same Chapman-Enskog expansion but the derivation is different.}
The leading order term of this distribution contains the Maxwellian of plasma~\eqref{eqn: maxwellian dist.}. 
Since the macroscopic quantities $u_p$ and $T_p$ in~\eqref{eqn: maxwellian dist.} are known and independent of the neutrals, the resulting fluid system contains more linear terms and is consequently expected to be computationally more efficient than the AFN model.
This fluid system will be used in the KDMC extension described in Sec.~\ref{sec: KDMC} and the hybrid model presented in Sec.~\ref{sec: hybrid model}.

\subsection{Diffusive scaling assumption}\label{sec: scaling}
The fluid model is valid in high-collisional regimes, in which we assume that the quantities of the simulation background scale with a small parameter $\varepsilon$, and the solution of \eqref{eqn: ss-eq}
can be expanded as 
\begin{equation}\label{eqn: phi expansion}
    \phi(x,v)= \phi_0(x,v) + \varepsilon \phi_1(x,v) + \varepsilon^2 \phi_2(x,v) + \cdots.
\end{equation}
In the so-called diffusive scaling \cite{maesHilbertExpansionBased2023a}, we assume that the velocity of an individual ion particle, \( v_p \sim \mathcal{O}(1/\varepsilon) \), is high, while the mean plasma velocity, \( u_p(x) \sim \mathcal{O}(1) \), is relatively low. By introducing the peculiar velocity $c_p(x)=v_p-u_p(x)$, the velocity variance of the plasma is defined as $\sigma_p^2(x) = \int c(x)^2 M_p(v\, | \, x)\, dv \sim \mathcal{O}(1/\varepsilon^2)$, and $\sigma_p^2(x)=T_p(x)/m$. 
In high-collisional regimes, neutral particles are close to the equilibrium of the plasma particles, so the velocity of a neutral particle $v$ has the same order of magnitude as $v_p$, and the velocity is decomposed as $v = c(x)+u_p(x)$, where $c(x)$ is the peculiar velocity of neutral particles. 
Moreover, the charge-exchange collision with the rate $R_{cx}(x)\sim\BO(1/\varepsilon^2)$ dominates in this regime, meaning that particles perform the charge-exchange collision frequently. Last, the ionization frequency is assumed to be low, i.e., $R_i(x)\sim\BO(1)$.  
Note that quantities with the subscript $p$, e.g., $v_p$, $u_p$, and $c_p$, are the quantities of plasma, and those without a subscript are the quantities of neutral particles.

\subsection{Fluid system--density equation}\label{sec: fluid system n}
To include the ionization term, we first add and subtract $R_iM_p(v)\intv \phi(v)dv'$ in \eqref{eqn: ss-eq}, which yields
\begin{equation}\label{eqn: ss-eq 2}
    v\partial_x \phi(v) = (R_{cx}+R_i)\left(M_p\left( v\right)\intv \phi(v')dv'-\phi(v)\right) - R_iM_p(v)\intv \phi(v)dv'+s(v),
\end{equation}
For notational simplicity, the spatial dependence is omitted if necessary. The procedure of the Hilbert expansion is widely discussed in literature (i.e., \cite{cercignaniMathematicalTheoryDilute1994a, harrisIntroductionTheoryBoltzmann2004a}), therefore, we only describe the main idea, and refer to~\cite{maesHilbertExpansionBased2023a} for the details.

Substituting the scaling assumption in Sec. \ref{sec: scaling} and the expansion \eqref{eqn: phi expansion} into \eqref{eqn: ss-eq 2}, we get a series of equations with different orders in the small parameter $\varepsilon$. 
The equation with the leading order term gives $\phi_0(x,v)=n(x,v)M_p(v\, | \, x)$. 
The equation with the first order term gives $\varepsilon \phi_1(x,v) = -1/(mR_t(x))\partial_x\left( c(x,v)n(x)M_p(v\, | \, x)\right)$ where $R_t(x)=R_{cx}(x)+R_i(x)$ the total collision rate, and $M_p(v\, | \, x)$ the plasma distribution Eq.~\eqref{eqn: maxwellian dist.}. Then, we obtain the approximation of the neutral distribution up to the first order 
\begin{equation}\label{eqn: phi appox}
    \phi(x,v)\approx n(x)M_p(v\, | \, x) - \frac{1}{mR_t(x)}\partial_x\left( c(x,v)n(x)M_p(v\, | \, x)\right).
\end{equation}
Next, the so-called solvability condition of the equation of the second-order term leads to a constraint on the neutral density $n(x)$, which reads
\begin{equation}\label{eqn: ss-density}
    \partial_x \left(u_p(x)n(x)\right)-\partial_x\left(\frac{1}{mR_t(x)}\partial_x \left(T_p(x)n(x)\right)\right)=R_r(x)n_p(x)-R_i(x)n(x).
\end{equation}
This is the density equation of the proposed fluid system. Again, macroscopic quantities with the subscript $p$ are the quantities of plasma, and those without a subscript are the quantities of neutral particles. 
Eq.~\eqref{eqn: ss-density} is a linear and steady-state equation.  
KDMC, as discussed in Sec. \ref{sec: KDMC}, requires the time-dependent density equation. To obtain the equation, we repeat the derivation above, starting from the time-dependent kinetic equation \eqref{eqn: t kinetic} with its initial condition $f(x,v,t=0)=R_r(x)M_p(v\, | \, x)n_p(x)$, which leads to the time-dependent density equation
\begin{equation}\label{eqn: t-density}
    \partial_tn(x,t) + \partial_x \left(u_p(x)n(x,t)\right)-\partial_x\left(\frac{1}{mR_t(x)}\partial_x \left(T_p(x)n(x,t)\right)\right)=-R_i(x)n(x,t),
\end{equation}
with the initial condition $n(x,t=0)=R_r(x)n_p(x)$.

Since the approximate neutral distribution \eqref{eqn: phi appox} is known, it can be substituted into the definition of the moments \eqref{eqn: moments}. This yields the expressions for the first and second moments,
\begin{equation} \label{eqn: moments approx}
m_1(x) =u_p(x)n(x) - \frac{1}{mR_t(x)}\partial_x\left(T_p(x)n(x)\right),  \quad m_2(x) = \frac{1}{2}\left(\frac{T_p(x)}{m}+u_p^2(x)\right)n(x)-\frac{1}{mR_t(x)}\partial_x\left(u_p(x)T_p(x)n(x)\right). 
\end{equation}
We observe that the neutral density $n(x)$ is the only unknown quantity, and it becomes available once the density equation \eqref{eqn: ss-density} is solved. 
Consequently, the QoIs defined in Eq.~\eqref{eqn: QoIs} can be obtained by solving only the density equation.
However, the approximation in Eq.~\eqref{eqn: moments approx} is only valid when the diffusive scaling assumptions are well satisfied. 
This is hard to satisfy in the plasma edge simulation, as will be shown in Sec.~\ref{sec: num}. 
Moreover, when realistic BCs are imposed, for instance, the reflective BC, the system deviates significantly from a Maxwellian distribution in the vicinity of the boundary.
As a result, the Hilbert expansion fails to capture these boundary effects and is only valid for describing the interior (bulk) solution.
To address these issues, we derive a more accurate fluid model by introducing additional momentum and energy equations, as done in the AFN model \cite{horstenDevelopmentAssessment2D2017}.

\subsection{Fluid system--momentum and energy equations}\label{sec: fluid system m and e}
We now extend the model with momentum and energy equations.
The purpose of the additional equations is not primarily to introduce a new stand-alone fluid model, but to provide an accurate fluid component to estimate those moments in the hybrid framework. We start from the steady-state linear kinetic equation \eqref{eqn: ss-eq}. For convenience of notation, we define 
\begin{equation}\label{eqn: coll op}
    g(\phi) =  R_{cx}\left(M_p\intv \phi dv'-\phi\right) - R_i \phi.
\end{equation}
That is, $g(\phi)$ contains the charge-exchange and the ionization terms of Eq. \eqref{eqn: ss-eq}. 
The dependence on $(x,v)$ is omitted for notational simplicity.
Next, multiplying Eq. \eqref{eqn: ss-eq} for $mv$ and $\frac{1}{2}mv^2$, and integrating over $v$, we obtain
\begin{align}\label{eqn: ss-mom}
    \partial_x\left(mnu^2+p\right) &= \intv mv g(\phi) dv + \intv mv sdv, \\ \label{eqn: ss-energy}
    \partial_x\left((\frac{1}{2}mu^2+\frac{1}{2}T)un+pu+q\right) &= \intv \frac{1}{2}mv^2g(\phi)dv + \intv \frac{1}{2}mv^2sdv,
\end{align}
where 
\begin{equation}\label{eqn: p & q}
    p=\intv m(u-v)^2\phi dv, \quad q=\intv \frac{1}{2}m(u-v)^3\phi dv.
\end{equation}
The integrals of $g(\phi)$ are
\begin{align}
    \intv mv g(\phi) dv &= R_{cx}\left(mu_p-mu\right)n - R_i m un,\\
    \intv \frac{1}{2}mv^2 g(\phi) dv &= R_{cx}\left[\left(\frac{1}{2}mu_p^2+\frac{1}{2}T_p\right)-\left(\frac{1}{2}mu^2+\frac{1}{2}T\right)\right]n - R_i \left(\frac{1}{2}mu^2+\frac{1}{2}T\right)n.
\end{align}
The integrals of the source term $s=R_rM_pn_p$ are
\begin{equation}
    \intv mv s dv = R_r m u_pn_p \quad \text{and} \quad \intv \frac{1}{2}mv^2sdv = R_r\left(\frac{1}{2}mu_p^2 + \frac{1}{2}T_p\right)n_p
\end{equation}
At this stage, the quantities $p$ and $q$ defined in \eqref{eqn: p & q} are unknown, since the distribution function $\phi$ is unknown. 
Consequently, the momentum equation \eqref{eqn: ss-mom} and the energy equation \eqref{eqn: ss-energy} are not closed. 
With the Chapman--Enskog framework, we could close \eqref{eqn: ss-mom} and \eqref{eqn: ss-energy} by approximating $\phi$ in \eqref{eqn: p & q} using the approximation obtained from the Hilbert expansion \eqref{eqn: phi appox}. The approximations are
\begin{equation}\label{eqn: p & q Euler}
    p\approx n T_p - \frac{3}{R_{t}m}nT_p\partial_xu_p, \quad \text{and} \quad q\approx - \frac{3}{2R_tm}\partial_x nT_p^2,
\end{equation}
in one-dimensional cases. Substituting $p$ and $q$ in \eqref{eqn: p & q Euler} into the momentum and energy equations \eqref{eqn: ss-mom} and \eqref{eqn: ss-energy}, we obtain the purely advection equations that lack the diffusion term (e.g., the heat conduction in the energy equation), since $q$ depends on $T_p$, not the unknown variable $T$.  
The resulting equations can be used for cases with periodic boundary conditions, but they cannot capture the boundary layer caused by the reflective BC in nuclear fusion cases, see the test case in Sec.~\ref{sec: num}.  
We thus need Navier-Stokes-type equations, as done in the AFN model~\cite{horstenDevelopmentAssessment2D2017}. To do so, we make the following modifications. 

First, in high-collisional regimes, we assume that the Maxwellian of neutral $M(v\, | \, x)$ is close to the Maxwellian of plasma $M_p(v\, | \, x)$  so that the $M_p(v\, | \, x)$ in the second term of the approximate $\phi$ \eqref{eqn: phi appox} is replaced by $M(v\, | \, x)$. Compared with the Maxwellian of plasma $M_p(v\, | \, x)$, the Maxwellian of neutrals reads 
\begin{equation}\label{eqn: maxwellian neutral}
    M(v\, | \, x)=M\left(v \mid u(x), T(x)\right) = \frac{\sqrt{m}}{\sqrt{2\pi T(x)}}\exp\left(\frac{|v-u(x)|^2}{2T(x)/m}\right),
\end{equation}
where the unknown functions $u(x)$ and $T(x)$ are respectively the macroscopic velocity and temperature of neutrals.
With this modification, $q$ will depend on $T$, instead of $T_p$.

Second, in the diffusive scaling assumption in Sec.~\ref{sec: scaling}, we assume $v = c(x) + u_p(x)\sim \BO(1/\varepsilon)$ and $u_p(x)\sim\BO(1)$. 
However, in plasma edge cases, the assumption is not always true, see the one-dimensional representative of the detached ITER case in Sec.~\ref{sec: num}. 
To compensate for the fact that $u_p(x)$ may not be $\BO(1)$ in some regimes but $u_p\sim v$, we replace the coefficient $c(x)$ in the second term of the approximation \eqref{eqn: phi appox} by $c(x)+u(x)$ as
\begin{equation}\label{eqn: phi appox ns}
    \phi(x,v)\approx n(x)M_p(v\, | \, x) - \frac{1}{mR_t(x)}\partial_x\left( (c(x)+u_p(x))n(x)M(v\, | \, x)\right) = n(x)M_p(v\, | \, x) - \frac{1}{mR_t(x)}v\partial_x\left( n(x)M(v\, | \, x)\right),
\end{equation}
where $v=c(x)+u(x)$ is taken out of the spatial derivative.
This modification does not violate the diffusive scaling assumption in the high-collision regimes, since $u_p(x)\sim\BO(1)$ is negligible compared with $c(x)$ in these regimes.

Last, the Chapman-Enskog method recovers a Navier-Stokes-type system if Eq.~\eqref{eqn: ss-eq} describes a self-collision model, in which the Maxwellian should be $M(v\, | \, x)$ instead of $M_p(v\, | \, x)$. In nuclear fusion cases, we assume the difference between $M(v\, | \, x)$ and $M_p(v\, | \, x)$ is small in the high-collisional regime \cite{horstenDevelopmentAssessment2D2017}, and we can neglect the difference. This allows us to substitute the neutral distribution approximation \eqref{eqn: phi appox ns} into the momentum equation \eqref{eqn: ss-mom} and the energy equation \eqref{eqn: ss-energy}, and subsequently apply the Chapman–Enskog method. Consequently, we obtain 
\todo{further linearization}
\begin{equation}\label{eqn: p & q NS}
    p \approx nT_p, \quad \text{and} \quad q \approx -\frac{3}{2R_tm}nT\partial_x T\approx-\frac{3}{2R_tm}nT_p\partial_x T,
\end{equation}
in one-dimensional cases, where the first temperature $T$ in $q$ is further approximated by $T_p$, the temperature of ions, to reduce the nonlinearity.
This further approximation follows the last assumption discussed above.

\todo{Explained here already that only density and energy eqs. are used. It was in the numerical experiment section}
\begin{hlblock}
In charge-exchange dominated regions, it is expected that the neutral drift velocity will remain small compared to the thermal velocity due to the decelerating friction force of ions flowing in the opposite direction~\cite{horstenComparisonFluidNeutral2016a}. 
In this case, the convective term $mnu^2$ in the momentum equation~\eqref{eqn: ss-mom} is neglected. Then, we do not calculate the neutral velocity $u$ by solving Eq.~\eqref{eqn: ss-mom}. 
Instead, we could simply calculate the velocity by $u=\Gamma/n$, where $\Gamma$ is the neutral particle flux. 
In our model, the flux is given by 
\begin{equation}\label{eqn: gamma}
    \Gamma = u_p(x)n(x) - \frac{1}{mR_t(x)}\partial_x\left(T_p(x)n(x)\right),
\end{equation}
which is, in fact, identical to the moment $m_1$ in Eq.~\eqref{eqn: moments approx} in the single-density model. 
Note that in two-dimensional cases, viscous effects become significant, and the full momentum equation must be solved to obtain $u$.
The argument is also true for the AFN model.
\end{hlblock}
Finally, our Navier-Stokes-type one-dimensional fluid system consists of the density equation \eqref{eqn: ss-density}, and the energy equation \eqref{eqn: ss-energy} with $p$ and $q$ in \eqref{eqn: p & q NS}.
\todo{Removed the statement about the neutral distribution approximation compared with the AFN model}

The numerical experiment in Sec.~\ref{sec: exp 1 num fluid model} shows that the approximate neutral distribution provides a good approximation at the boundary with the absorbing BC.
In addition, unlike the AFN model, the density and energy equations of the proposed model are linear and do not couple with each other.
Therefore, the computational cost of solving the proposed fluid model is lower. 
\todo{In the new fluid model, we claim the model has more linear terms. We don't say it is a linear model. Having more linear terms is already faster.}
Although our model contains more linear terms, in the numerical experiment in Sec.~\ref{sec: exp 1 num fluid model}, we will see that the accuracy of these two fluid models has no obvious difference.
The derivation of the fluid system in two- and three-dimensional cases is left as future work.
\section{Kinetic-Diffusion Monte Carlo}\label{sec: KDMC}
In this section, we explain the kinetic-diffusion Monte Carlo (KDMC) particle method proposed in \cite{mortierKineticDiffusionAsymptoticPreservingMonte2022} for simulating the neutral particle trajectories governed by the time-dependent kinetic equation \eqref{eqn: t kinetic}
and we extend KDMC to account for ionization collisions. 
KDMC is an asymptotic-preserving method that adaptively combines kinetic steps that mimic physical particle movements, with diffusive steps that represent the accumulated effect of a large number of kinetic steps. The former is called the kinetic (MC) simulation, and the latter is referred to as the diffusion simulation.

The role of KDMC in the present work is twofold: it provides efficient trajectory generation, and it
induces the particle-level decomposition on which the hybrid formulation is built.
Next, we begin in Sec.~\ref{sec: kmc} with the kinetic MC simulation, followed by the diffusion simulation derived from the time-dependent density equation~\eqref{eqn: t-density}, in Sec.~\ref{sec: dmc}. 
Finally, in Sec.~\ref{sec: kd}, we present KDMC.

\subsection{Kinetic simulation}\label{sec: kmc}
To simulate the discretized trajectories by kinetic MC, we consider the $i$-th particle with $i=1, 2, \cdots, I$ and the weight $w_i$, where $I$ is the number of particles. 
The weight indicates the fraction of the total mass that the particle represents for the system. 
In particular, if the system has the initial density $s(x,v)=R_r(x)M_p(v\mid x)n_p(x)$ as explained in Sec.~\ref{sec: t kinetic eq}, the weight is then $w_i = \int_\Omega\intv s(x,v)dvdx/I$ for all $i$ with $x\in\Omega$ the simulation domain. 
Next, the simulation of this particle starts from the time $t=0$ at the position $x^0_i$ sampled from $R_r(x)n_p(x)$ with the velocity $v^0_i$ sampled from $M_p(v\mid x_i^0)$. 
After a flight time $\tau$ sampled as
$ \int_0^\tau R_t(x_i^0 + v_i^0s)ds=\mathcal{E} \sim \mathrm{Exp}(1),$
a collision event occurs. \todo{The distribution of $\tau$ was wrong. I modified it} 
Here, $R_t(x^0_i)=R_{cx}(x^0_i)+R_i(x^0_i)$ the total collision rate at $x_i^0$. The particle position is then updated as 
\begin{equation}\label{eqn: a kinetic step}
    x^1_i = x^0_i + \tau v^0_i,
\end{equation} 
With probability $R_i(x^1_i)/R_t(x^1_i)$, the collision type is ionization, and the simulation of this particle stops since it is ionized. Otherwise, namely, with probability $R_{cx}(x^1_i)/R_t(x^1_i)$, the collision type is charge-exchange,  
and the particle is assigned a new velocity $v^1_i$ sampled from the Maxwellian distribution $M_p(v|x^1_i)$ given in \eqref{eqn: maxwellian dist.}.
The former corresponds to an ionization process, while the latter represents a velocity-jump process.
The movement \eqref{eqn: a kinetic step} followed by the ionization process or the velocity jump process is called a kinetic step.
Repeating the kinetic step until the particle is ionized, we obtain its trajectory denoted as $\{(x_i^j, v_i^j)\}_{j=0}^{J_i}$, where $J_i$ is the total number of states of the $i$-th particle. The trajectories of the particle system are then constructed by applying the same procedure to all particles.

\subsection{Diffusion simulation}\label{sec: dmc}
In this section, we derive the diffusive movement of particles used in KDMC, starting from the time-dependent density equation \eqref{eqn: t-density}. Building upon the original KDMC method \cite{mortierKineticDiffusionAsymptoticPreservingMonte2022}, we incorporate the ionization process in this movement with the operator splitting method~\cite{macnamaraOperatorSplitting2016c}. 

\todo{Before, I first rewrote the eq, and performed operator splitting. Now, first splitting and rewriting. We rewrite it into the Fokker-Planck form.}
Applying the operator splitting method~\cite{macnamaraOperatorSplitting2016}, the density equation~\eqref{eqn: t-density} is approximated by sequentially solving  
\begin{align}\label{eqn: op 1}
    &\partial_tn(x,t) + \partial_x \left(u_p(x)n(x,t)\right)-\partial_x\left(\frac{1}{mR_t(x)}\partial_x \left(T_p(x)n(x,t)\right)\right) =0, \\ \label{eqn: op 2}
    &\partial_t n(x,t) = -R_i(x)n(x,t).
\end{align} 
For Eq.~\eqref{eqn: op 1},  we add and subtract the term  $\partial_x\left(\partial_x(\frac{1}{m R_t(x)})T_p(x) n(x,t)\right)$ on the left-hand side, and it becomes
\begin{equation}\label{eqn: fp sink}
    \partial_t n(x,t) + \partial_x\left(\left(u_p(x)+\partial_x\left(\frac{1}{mR_t(x)}\right)T_p(x)\right)n(x,t)\right) - \partial_{xx}\left(\frac{T_p(x)}{mR_t(x)}n(x,t)\right) = 0,
\end{equation}
a Fokker-Planck equation 
\cite{lapeyreIntroductionMonteCarloMethods2003b}, whose Ito stochastic differential equation (SDE) is
\begin{equation}\label{eqn: SDE of FK}
    dX_t = \left(u_p(X_t)+\partial_x\left(\frac{1}{mR_t(X_t)}\right)T_p(X_t)\right)dt + \sqrt{2\frac{T_p(X_t)}{mR_t(X_t)}}dW_t.
\end{equation}
where $X_t$ is the stochastic position process of particles and $W_t$ is a Brownian motion. Let
$$
A(x) = u_p(x)+\partial_x\left(\frac{1}{mR_t(x)}\right)T_p(x), \quad \text{ and } \quad D(x) = \frac{T_p(x)}{mR_t(x)}. 
$$
Applying the Euler-Maruyama approximation,  the SDE~\eqref{eqn: SDE of FK} indicates that the new position that the $i$-th particle travels a small time step of size $\theta$ from $x_i^j$ is 
\begin{equation}\label{eqn: SDE update}
    x_i^{j'} = x_i^j + A(x_i^j)\theta + \sqrt{2D(x_i^j)\theta}\xi,
\end{equation}
with $\xi\sim\mathcal{N}(0,1)$. 
The ionization part, i.e., Eq.~\eqref{eqn: op 2}, has the approximate solution
\begin{equation}\label{eqn: SDE ionize}
    n(x,t+\theta)\approx \exp(-R_i(x)\theta) n(x,t)
\end{equation}
with a small time step of size $\theta$. At the particle level, Eq.~\eqref{eqn: SDE ionize} means the $i$-th particle is ionized with probability $1-\exp(-R_i(x_i^j)\theta)$ traveling from $x_i^k$ after a time step $\theta$  \cite{dimarcoAsymptoticPreservingMonteCarlo2018c}. The simulation of this particle stops when it is ionized.

In conclusion, in the diffusion simulation, the $i$-th particle first flies from $x_i^j$ to $x_i^{j'}$ with velocity $v_i^j$ over a user-defined time step of size $\theta$. After this time interval, the particle is ionized (and thus removed from the simulation) with probability $1-\exp(-R_i(x_i^j)\theta)$ or survives with probability  $\exp(-R_i(x_i^j)\theta)$.
The movement \eqref{eqn: SDE update} with the ionization process given by~\eqref{eqn: SDE ionize} is called a diffusive step. 
When the scaling parameter $\varepsilon\rightarrow 0$, the true particle movement following Eq. \eqref{eqn: t kinetic} converges to this step. 

\subsection{Kinetic-diffusion simulation}\label{sec: kd}
The kinetic simulation described in Sec.~\ref{sec: kmc} is unbiased~\cite{luxMonteCarloParticle2018a} but computationally expensive in highly collisional regimes. 
In contrast, the diffusion simulation introduced in Sec.~\ref{sec: dmc} offers a significant reduction in computational cost in these regimes, with only a minor loss in accuracy.
However, in low-collisional regimes, the density equation~\eqref{eqn: t-density} and thus the diffusion simulation no longer provides a good approximation, whereas the kinetic simulation is both valid and efficient.
In the intermediate regime, one must choose between accuracy and efficiency. 

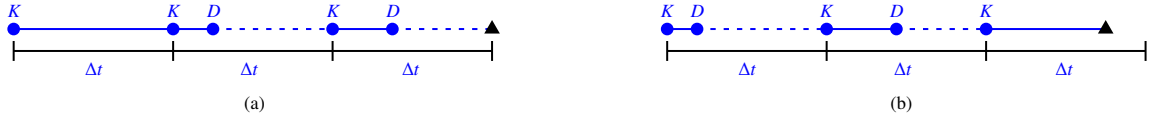
\begin{figure}[h]
	\centering

    \begin{subfigure}{0.48\textwidth}
    \centering
	\tikzset{every picture/.style={line width=0.75pt}} 
	\begin{tikzpicture}[x=0.75pt,y=0.75pt,yscale=-1,xscale=1]
		\draw[color=KULeuvenLichtblauw] (80,120) -- (160,120);
		\filldraw[color=KULeuvenLichtblauw] (80,120) circle (2pt);
		
		\filldraw[color=KULeuvenLichtblauw] (160,120) circle (2pt);
		
		\draw[color=KULeuvenLichtblauw] (160,120) -- (180,120);
		\filldraw[color=KULeuvenLichtblauw] (180,120) circle (2pt);
		
		\draw[color=KULeuvenLichtblauw,dash pattern={on 2pt off 3pt}] (180,120) -- (240,120);
		\filldraw[color=KULeuvenLichtblauw] (240,120) circle (2pt);

		\draw[color=KULeuvenLichtblauw,dash pattern={on 2pt off 3pt}] (270,120) -- (320,120);
		\filldraw[color=KULeuvenLichtblauw] (270,120) circle (2pt);
		
		\draw[color=KULeuvenLichtblauw] (240,120) -- (270,120);
		
		\filldraw[color=black] (320,120) node[regular polygon, regular polygon sides=3, 
         minimum size=4pt, inner sep=1.2pt, fill=black] {};

		\draw (80,131) -- (320,131);
		
		\draw (80,126) -- (80,136);
		\draw (160,126) -- (160,136);
		\draw (240,126) -- (240,136);
		\draw (320,126) -- (320,136);
		
		\draw (273,135) node [anchor=north west][inner sep=1pt, color=KULeuvenLichtblauw] {\fontsize{7}{8}\selectfont $\Delta t$};
		\draw (192,135) node [anchor=north west][inner sep=1pt, color=KULeuvenLichtblauw] {\fontsize{7}{8}\selectfont $\Delta t$};
		\draw (114,135) node [anchor=north west][inner sep=1pt, color=KULeuvenLichtblauw] {\fontsize{7}{8}\selectfont $\Delta t$};

	\draw (75,106) node [anchor=north west][inner sep=1pt, color=KULeuvenLichtblauw] {\scriptsize$K$};
	\draw (155,106) node [anchor=north west][inner sep=1pt, color=KULeuvenLichtblauw] {\scriptsize$K$};
	\draw (175,106) node [anchor=north west][inner sep=1pt, color=KULeuvenLichtblauw] {\scriptsize$D$};				
	\draw (235,106) node [anchor=north west][inner sep=1pt, color=KULeuvenLichtblauw] {\scriptsize$K$};
	\draw (265,106) node [anchor=north west][inner sep=1pt, color=KULeuvenLichtblauw] {\scriptsize$D$};
    
	\end{tikzpicture}
	\label{fig: KDMC-a}
    \caption{}
    \end{subfigure}
    \hfill
    \begin{subfigure}{0.48\textwidth}
	\centering
    \tikzset{every picture/.style={line width=0.75pt}} 
	\begin{tikzpicture}[x=0.75pt,y=0.75pt,yscale=-1,xscale=1]
	\draw[color=KULeuvenLichtblauw] (80,180) -- (95,180);
	\filldraw[color=KULeuvenLichtblauw] (80,180) circle (2pt);
	\filldraw[color=KULeuvenLichtblauw] (95,180) circle (2pt);
	
	\draw[color=KULeuvenLichtblauw,dash pattern={on 2pt off 3pt}] (95,180) -- (160,180);
	\filldraw[color=KULeuvenLichtblauw] (160,180) circle (2pt);
	
	\draw[color=KULeuvenLichtblauw] (160,180) -- (195,180);
	\filldraw[color=KULeuvenLichtblauw] (195,180) circle (2pt);
	
	\draw[color=KULeuvenLichtblauw,dash pattern={on 2pt off 3pt}] (195,180) -- (240,180);
	\filldraw[color=KULeuvenLichtblauw] (240,180) circle (2pt);

	\draw[color=KULeuvenLichtblauw] (240,180) -- (300,180);
	
	\filldraw[color=black] (300,180) node[regular polygon, regular polygon sides=3, 
         minimum size=4pt, inner sep=1.2pt, fill=black] {};
	
	\draw (80,191) -- (320,191);
	
	\draw (80,186) -- (80,196);
	\draw (160,186) -- (160,196);
	\draw (240,186) -- (240,196);
	\draw (320,186) -- (320,196);
	
	\draw (273,195) node [anchor=north west][inner sep=1pt, color=KULeuvenLichtblauw] {\fontsize{7}{8}\selectfont $\Delta t$};
	\draw (192,195) node [anchor=north west][inner sep=1pt, color=KULeuvenLichtblauw] {\fontsize{7}{8}\selectfont $\Delta t$};
	\draw (114,195) node [anchor=north west][inner sep=1pt, color=KULeuvenLichtblauw] {\fontsize{7}{8}\selectfont $\Delta t$};
	
	\draw (75,166) node [anchor=north west][inner sep=1pt, color=KULeuvenLichtblauw] {\scriptsize\textit{K}};
	\draw (90,166) node [anchor=north west][inner sep=1pt, color=KULeuvenLichtblauw] {\scriptsize\textit{D}};
	\draw (155,166) node [anchor=north west][inner sep=1pt, color=KULeuvenLichtblauw] {\scriptsize\textit{K}};
	\draw (190,166) node [anchor=north west][inner sep=1pt, color=KULeuvenLichtblauw] {\scriptsize\textit{D}};
	\draw (235,166) node [anchor=north west][inner sep=1pt, color=KULeuvenLichtblauw] {\scriptsize\textit{K}};

	\end{tikzpicture}
    \caption{}
	\label{fig: KDMC-b}
    \end{subfigure}
    \caption{Illustration of KDMC. (a) In the first time step of size $\dt$, the particle is in a low-collisional regime where only the kinetic (K) step is performed. In the second time step, the particle is in a high-collision regime where the kinetic step only lasts a short time, and the diffusive (D) step dominates. In the third time step, the particle is ionized after the diffusive step, as indicated by the black triangle. (b) In the third time step, the particle is ionized in the kinetic step, as indicated by the black triangle.}
    \label{fig: KDMC}
\end{figure}

These issues are handled in \cite{mortierKineticDiffusionAsymptoticPreservingMonte2022} by combining the advantages of the kinetic and the diffusion simulations. 
Specifically, we fix a time step of size $\dt$ and consider the $i$-th particle. 
Within the time interval, the particle first performs a kinetic step with time $\tau$ described in Sec.~\ref{sec: kmc}. 
If $\tau>\dt$, no collision occurs and no diffusive step is performed during this time step. If $\tau<\dt$, a collision executes at time $\tau$. 
If the collision is of an ionization type, the particle is removed from the simulation. 
If it is of charge-exchange type, the particle is assigned a new velocity as described in Sec.~\ref{sec: kmc}, and then performs a diffusive step described in Sec.~\ref{sec: dmc} over the remaining time $\theta = \dt -\tau$.
If the particle survives in the diffusive step (i.e., is not ionized), the procedure is repeated in the next time step of size $\dt$ until ionization occurs.
The kinetic and diffusive steps in different situations are illustrated in Fig.~\ref{fig: KDMC}.

Simulating the $i$-th particle from $t=0$ until ionization, and denoting $J_i$ the time step during which ionization occurs of the $i$-th particle, the trajectory of this particle can then be written as
\begin{equation}\label{eqn: trajectory}
    \{(x_i^0, v_i^0), ({x_i^0}', {v_i^0}'), \cdots, (x_i^{J_i}, v_i^{J_i}), (x_i^{J_i'}, v_i^{J_i'}) \} \quad \text{or} \quad \{(x_i^0, v_i^0), ({x_i^0}', {v_i^0}'),  \cdots, (x_i^{J_i}, v_i^{J_i}) \},
\end{equation}
depending on whether the particle is ionized in the $J_i$-th kinetic step or the $J_i$-th diffusive step, respectively.
The state $({x_i^{j}}, {v_i^{j}})$ with the index $j$ is the starting state of the $j$-th kinetic step, 
and the state $({x_i^{j}}', {v_i^{j}}')$ with the index $j'$ is that of the $j$-th diffusive step, where $j=0,1, \ldots, J_i$. 
The weight is $w_i = \int n(x,t=0)dx/I$  for all $i$ as stated in Sec. \ref{sec: kmc}. 

Intuitively, when the collision rate is high, i.e., $\tau \ll \dt$ on average, the particle will first perform a short kinetic step and then a dominant diffusive step for the rest of the time step $\dt$. 
On the contrary, when the collision rate is low, i.e., $\tau \gg \dt$ on average, particles are primarily simulated by the kinetic step, and the influence of the inexact diffusive step is insignificant.  In both situations, at most two flights happen per time step of size $\dt$.

\section{Hybrid model}\label{sec: hybrid model}
In this section, we derive the proposed hybrid model for solving the steady-state kinetic equation \eqref{eqn: ss-eq} 
by decomposing its solution (neutral distribution) into a kinetic part and a fluid part. Based on the KDMC simulation scheme presented in Sec.~\ref{sec: KDMC}, the decomposition is automatic and asymptotic-preserving. 
Afterwards, the QoIs defined in Sec.~\ref{sec: QoIs} contributed by the kinetic part are estimated by an MC estimator. 
The QoIs contributed by the fluid part are estimated by reconstructing and solving the steady-state fluid system proposed in Sec.~\ref{sec: fluid system}.

For better clarity, in the following, we first derive the hybrid model based on the particle trajectories generated by the kinetic MC simulation, in Sec.~\ref{sec: hybrid kmc}. 
We then describe how KDMC is applied in Sec.~\ref{sec: hybrid model kdmc}. 
The derivation is based on the mathematical connection between the time-dependent particle simulation and the steady-state problem, similar to the derivation in Sec.~\ref{sec: t kinetic eq}. But here, the connection is first applied to each time step and then to the whole simulation time.
Finally, in Sec.~\ref{sec: hybrid model comparison}, we compare our model and the scheme proposed in~\cite{mortierEstimationPostprocessingStep2022} that estimates the QoIs of the fluid part by solving a time-dependent equation.

\subsection{Hybrid model based on kinetic MC}\label{sec: hybrid kmc}
With the collision operator $g(f)$ defined in~\eqref{eqn: coll op}, the linear steady-state \eqref{eqn: ss-eq} and time-dependent \eqref{eqn: t kinetic} kinetic equations are then written as
\begin{equation}\label{eqn: ss-eq short}
	v\partial_x \phi(x,v) = g(\phi(x,v)) + s(x,v),
\end{equation}
and
\begin{equation}\label{eqn: t-eq short}
	\partial_t f(x,v,t) + v\partial_x f(x,v,t) = g(f(x,v,t)), \quad \text{with} \quad f(x,v,t=0)=s(x,v),
\end{equation}
respectively, where the steady-state distribution $\phi(x,v) = \int_0^{\infty} f(x,v,t)dt$, as stated in Sec.~\ref{sec: t kinetic eq}. 

To derive the hybrid model based on kinetic MC, we fix a time step of size $\dt$, as in KDMC, but let the particle execute kinetic steps all the time, without diffusive steps. 
Within a time step of size $\dt$, the trajectory of the $i$-th particle is denoted by the sequence of particle states
\begin{equation}\label{eqn: mc trajectory} 
\{(x_{i,k}^0, v_{i,k}^0), \ldots, (x_{i,k}^{\nu-1}, v_{i,k}^{\nu-1}), (x_{i,f}^{\nu}, v_{i,f}^{\nu}), \ldots, (x_{i,f}^{N-1}, v_{i,f}^{N-1})\},
\end{equation}
where $N$ is the total number of collisions during this time step, varying from particle to particle.
The second subscript $k$ or $f$ labels the particle states into the kinetic state and the fluid state, respectively.
In particular, the first $\nu$ states are the kinetic states, and the remaining $N-\nu$ states are the fluid states. Note that $\nu$ could be different for each particle, depending on the simulation scheme.
By labeling the particle states as kinetic or fluid, we use kinetic states in the kinetic part of the hybrid model and fluid states in the fluid part.
Note that we always know the state of a particle at time $j\dt$ in the kinetic MC simulation, such that the particle has a state exactly at $j\dt$ for $j=0,1,\ldots, \infty$.

In the following, we consider $\nu=2$, i.e., only one kinetic step within a time step as in KDMC, and the first two states are used in the kinetic part of the hybrid model, 
for instance, with the track-length estimator~\cite{luxMonteCarloParticle2018a}. 

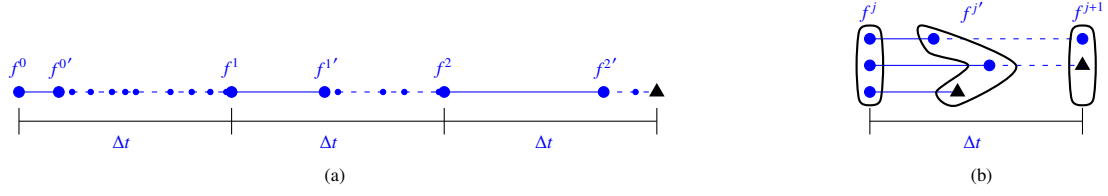
\begin{figure}
	\centering
    \begin{subfigure}{0.63\textwidth}
    \centering
    \begin{tikzpicture}[x=1pt,y=1pt,yscale=-1,xscale=1]
		\tikzset{every picture/.style={line width=0.75pt}}
		\draw[color=KULeuvenLichtblauw] (80,120) -- (95,120);
		\filldraw[color=KULeuvenLichtblauw] (80,120) circle (2pt);
		\filldraw[color=KULeuvenLichtblauw] (95,120) circle (2pt);
		
		\draw[color=KULeuvenLichtblauw,dash pattern={on 2pt off 3pt}] (95,120) -- (160,120);
        \filldraw[color=KULeuvenLichtblauw] (100,120) circle (1pt);
        \filldraw[color=KULeuvenLichtblauw] (107,120) circle (1pt);
        \filldraw[color=KULeuvenLichtblauw] (115,120) circle (1pt);
        \filldraw[color=KULeuvenLichtblauw] (120,120) circle (1pt);
        \filldraw[color=KULeuvenLichtblauw] (124,120) circle (1pt);
        \filldraw[color=KULeuvenLichtblauw] (137,120) circle (1pt);
        \filldraw[color=KULeuvenLichtblauw] (145,120) circle (1pt);
        \filldraw[color=KULeuvenLichtblauw] (152,120) circle (1pt);
        \filldraw[color=KULeuvenLichtblauw] (158,120) circle (1pt);
		\filldraw[color=KULeuvenLichtblauw] (160,120) circle (2pt);
		
		\draw[color=KULeuvenLichtblauw] (160,120) -- (195,120);
		\filldraw[color=KULeuvenLichtblauw] (195,120) circle (2pt);
		\filldraw[color=KULeuvenLichtblauw] (200,120) circle (1pt);
        \filldraw[color=KULeuvenLichtblauw] (217,120) circle (1pt);
        \filldraw[color=KULeuvenLichtblauw] (225,120) circle (1pt);
        \filldraw[color=KULeuvenLichtblauw] (238,120) circle (1pt);
		\draw[color=KULeuvenLichtblauw,dash pattern={on 2pt off 3pt}] (195,120) -- (240,120);
		\filldraw[color=KULeuvenLichtblauw] (240,120) circle (2pt);

		\draw[color=KULeuvenLichtblauw,dash pattern={on 2pt off 3pt}] (300,120) -- (320,120);
		\filldraw[color=KULeuvenLichtblauw] (300,120) circle (2pt);
		\filldraw[color=KULeuvenLichtblauw] (312,120) circle (1pt);
		\draw[color=KULeuvenLichtblauw] (240,120) -- (300,120);
		
		\filldraw[color=black] (320,120) node[regular polygon, regular polygon sides=3, 
         minimum size=4pt, inner sep=1.2pt, fill=black] {};
		
		\draw (80,131) -- (320,131);
		
		\draw (80,126) -- (80,136);
		\draw (160,126) -- (160,136);
		\draw (240,126) -- (240,136);
		\draw (320,126) -- (320,136);
		
		


		\draw (273,135) node [anchor=north west][inner sep=1pt, color=KULeuvenLichtblauw] {\fontsize{7}{8}\selectfont $\Delta t$};
		\draw (192,135) node [anchor=north west][inner sep=1pt, color=KULeuvenLichtblauw] {\fontsize{7}{8}\selectfont $\Delta t$};
		\draw (114,135) node [anchor=north west][inner sep=1pt, color=KULeuvenLichtblauw] {\fontsize{7}{8}\selectfont $\Delta t$};

		\draw (75,106) node [anchor=north west][inner sep=1pt, color=KULeuvenLichtblauw] {\scriptsize${f^0}$};
		\draw (90,106) node [anchor=north west][inner sep=1pt, color=KULeuvenLichtblauw] {\scriptsize${f^0}'$};
		\draw (155,106) node [anchor=north west][inner sep=1pt, color=KULeuvenLichtblauw] {\scriptsize${f^1}$};
		\draw (190,106) node [anchor=north west][inner sep=1pt, color=KULeuvenLichtblauw] {\scriptsize${f^1}'$};				
		\draw (235,106) node [anchor=north west][inner sep=1pt, color=KULeuvenLichtblauw] {\scriptsize${f^2}$};
		\draw (295,106) node [anchor=north west][inner sep=1pt, color=KULeuvenLichtblauw] {\scriptsize${f^2}'$};

	\end{tikzpicture}
	\caption{  } 
	\label{fig: ss_to_t}
    \end{subfigure}
    \hfill
    \begin{subfigure}{0.33\textwidth}
    \centering
	\begin{tikzpicture}[x=1pt,y=1pt,yscale=-1,xscale=1]
    \tikzset{every picture/.style={line width=0.75pt}}
		
		\draw[color=KULeuvenLichtblauw] (80,120) -- (113,120);
		\filldraw[color=KULeuvenLichtblauw] (80,120) circle (2pt);
		
        \filldraw[color=black] (113,120) node[regular polygon, regular polygon sides=3, 
         minimum size=4pt, inner sep=1.2pt, fill=black] {};

        \draw[color=KULeuvenLichtblauw] (80,110) -- (125,110);
		\filldraw[color=KULeuvenLichtblauw] (80,110) circle (2pt);
		\filldraw[color=KULeuvenLichtblauw] (125,110) circle (2pt);
		\draw[color=KULeuvenLichtblauw,dash pattern={on 2pt off 3pt}] (125,110) -- (160,110);
        \filldraw[color=black] (160,110) node[regular polygon, regular polygon sides=3, 
         minimum size=4pt, inner sep=1.2pt, fill=black] {};

        \draw[color=KULeuvenLichtblauw] (80,100) -- (104,100);
		\filldraw[color=KULeuvenLichtblauw] (80,100) circle (2pt);
		\filldraw[color=KULeuvenLichtblauw] (104,100) circle (2pt);
		\draw[color=KULeuvenLichtblauw,dash pattern={on 2pt off 3pt}] (104,100) -- (160,100);
		\filldraw[color=KULeuvenLichtblauw] (160,100) circle (2pt);

        \draw[thick]
        plot[smooth cycle] coordinates {
        (100,95)
        (135,110)
        (113,125)
        (105,120)
        (117, 110)
        (102,105)
        };
        
        \draw[thick]
        plot[smooth cycle] coordinates {
        (80,95)
        (84, 97)
        (85, 110)
        (84, 123)
        (80,125)
        (76, 123)
        (75, 110)
        (76, 97)
        };

        \draw[thick]
        plot[smooth cycle] coordinates {
        (80+80,95)
        (84+80, 97)
        (85+80, 110)
        (84+80, 123)
        (80+80,125)
        (76+80, 123)
        (75+80, 110)
        (76+80, 97)
        };
        
		\draw (80,131) -- (160,131);
		
		\draw (80,126) -- (80,136);
		\draw (160,126) -- (160,136);
		
		

		\draw (75,85) node [anchor=north west, inner sep=1pt, color=KULeuvenLichtblauw]
		{\scriptsize$f^j$};

        \draw (113,85) node [anchor=north west, inner sep=1pt, color=KULeuvenLichtblauw]
		{\scriptsize${f^j}'$};

        \draw (155,85) node [anchor=north west, inner sep=1pt, color=KULeuvenLichtblauw]
		{\scriptsize$f^{j+1}$};


		\draw (114,135) node [anchor=north west][inner sep=1pt, color=KULeuvenLichtblauw] {\fontsize{7}{8}\selectfont $\Delta t$};


	\end{tikzpicture}
	\caption{} 
	\label{fig: def f}
    \end{subfigure}
    \caption{Kinetic MC simulation with a fixed time step $\Delta t$, assuming one collision in the kinetic part, as in KDMC. A black triangle denotes the ionization of a particle. (a) Solid lines indicate the kinetic stages, and dashed lines indicate the fluid stages.
    There may be more than one (charge-exchange) collision (denoted by small dots) within a fluid stage.
    (b) Illustration of the distributions $f^j$ and ${f^j}'$. The samples of ${f^j}'$ are generally at different time instants, since particles collide at different times. }
\end{figure}

To clear the time intervals that the kinetic and fluid parts of the proposed hybrid model apply, we define
\begin{itemize}
    \item{Kinetic stage:} the time interval $(j\dt, j\dt+\tau]$, if $\tau<\dt$, where $\tau$ is the collision time from $j\dt$ to the first collision; Otherwise, $(j\dt, (j+1)\dt]$.
    \item{Fluid stage:} the time interval $(j\dt+\tau, (j+1)\dt]$, if $\tau<\dt$; Otherwise, $\emptyset$, i.e., no fluid stage.
\end{itemize}
The kinetic and fluid stages, illustrated in Fig.~\ref{fig: ss_to_t}, vary from particle to particle, since their collision times are random.
In Sec.~\ref{sec: hybrid model kdmc}, the fluid states in a fluid stage will be replaced by a single diffusive step as KDMC. The error of replacing the collisions in the fluid stage with a single diffusive step is analyzed in~\cite{tangAnalysisKineticdiffusionMonte2025}. It is called the KDMC simulation error.

With the kinetic and fluid states, we decompose the time-dependent distribution $f(x,v,t)$ as
\begin{equation}
    f(x,v,t)=f_k(x,v,t)+f_f(x,v,t)
\end{equation}
where $f_k(x,v,t)$ and $f_f(x,v,t)$ are the distribution of the kinetic states and the fluid states, respectively, and thus the kinetic and fluid parts of the distribution $f(x,v,t)$.
Correspondingly, we have
\begin{equation*}
	\phi_k(x,v) = \int_0^{\infty} f_k(x,v,t)dt, \quad \text{and} \quad \phi_f(x,v) = \int_0^{\infty} f_f(x,v,t)dt,
\end{equation*}
the steady-state distributions contributed by the states of the kinetic part and the states of the fluid part, respectively.
Now, consider the $j$-th time interval of size $\Delta t$, $t\in[j\dt, (j+1)\dt]$.
The steady-state neutral distribution integrated from the $j$-th time interval is
\begin{equation}\label{eqn: phi j}
	\phi^j(x,v) = \int_{j\dt}^{(j+1)\dt}f(x,v,t)dt=\int_{j\dt}^{(j+1)\dt}f_k(x,v,t)dt + \int_{j\dt}^{(j+1)\dt}f_f(x,v,t)dt=\phi_k^j(x,v) + \phi_f^j(x,v),
\end{equation}
where $\phi^j_k$ and $\phi^j_f$ are the kinetic and fluid part of $\phi^j$.
We then denote
\begin{itemize}
    \item $f^j(x,v)=f(x,v,j\dt)$ the time-dependent neutral distribution at time $t=j\dt$.
    \item ${f^j}'(x,v)$ the distribution that particles just finish their first collision after the time $t=j\dt$. If the collision time of a particle $\tau>\dt$, the particle has the state at $(j+1)\dt$. In other words, ${f^j}'(x,v)$ is the distribution at the beginning of the fluid stage.
\end{itemize}
The two notations are illustrated in Fig.~\ref{fig: def f}.
Note that, if no particle collides within the $j$-th time step, ${f^j}'=f^{j+1}$. 
Next, integrating the time-dependent equation \eqref{eqn: t-eq short} over the $j$-th time interval, we have
\begin{equation}\label{eqn: t-eq in idt}
	f^{j+1}(x,v) - f^j(x,v) + v\partial_x \left(\int_{j\dt}^{(j+1)\dt}f(x,v,t)dt \right) = g\left(\int_{j\dt}^{(j+1)\dt}f(x,v,t)dt\right)
\end{equation}	
Adding and subtracting ${f^j}'(x,v)$ on the left-hand side and with $f=f_k+f_f$, Eq.~\eqref{eqn: t-eq in idt} becomes
\begin{equation}
	\left(f^{j+1} -{f^j}' +{f^j}' - f^j\right)(x,v)+ v\partial_x \left( \int_{j\dt}^{(j+1)\dt} f_k(x,v,t) + f_f(x,v,t) dt \right) = g\left(\int_{j\dt}^{(j+1)\dt} f_k(x,v,t) + f_f(x,v,t) dt\right),
\end{equation}	
Decomposing the above equation into the kinetic part and fluid part and with Eq.~\eqref{eqn: phi j}, we get 
\begin{align}\label{eqn: i-k part eq}
	v\partial_x \phi^j_k(x,v) &= g\left(\phi^j_k(x,v)\right) + f^j(x,v) - {f^j}'(x,v), \\ \label{eqn: i-f part eq}
	v\partial_x \phi^j_f(x,v) &= g\left(\phi^j_f(x,v)\right) + {f^j}'(x,v) - f^{j+1}(x,v).
\end{align}
Finally, adding all time intervals with $j=0,1,\ldots,\infty$ together, the kinetic and fluid part Eqs.~\eqref{eqn: i-k part eq} and \eqref{eqn: i-f part eq} become
\begin{align}\label{eqn: k part eq}
	v\partial_x \phi_k(x,v) &= g\left(\phi_k(x,v)\right) + s_k(x,v), \quad \text{with} \quad s_k(x,v)=\sum_{j=0}^\infty\left(f^j(x,v) - {f^j}'(x,v)\right), \\ \label{eqn: f part eq}
	v\partial_x \phi_f(x,v) &= g\left(\phi_f(x,v)\right) + s_f(x,v), \quad \text{with} \quad  s_f(x,v)=\sum_{j=0}^\infty\left({f^j}'(x,v) - f^{j+1}(x,v)\right),
\end{align}
respectively, where $\phi_k(x,v)=\sum_{i=0}^\infty\phi_k^j(x,v)$ and $\phi_f(x,v)=\sum_{j=0}^\infty\phi_f^j(x,v)$. 

Eqs.~\eqref{eqn: k part eq} and \eqref{eqn: f part eq} bridge the time-dependent simulation and the steady-state model \eqref{eqn: ss-eq short} with a fixed time step of size $\dt$. 
The original steady-state equation \eqref{eqn: ss-eq short} is divided into a kinetic and a fluid part, where the corresponding source terms $s_k$ and $s_f$ are the sum of the difference between $f^j$ and ${f^j}'$.
It means, when estimating the QoIs (defined in Eq.\eqref{eqn: QoIs}) contributed by the kinetic part using the MC estimator, like the track-length estimator \cite{luxMonteCarloParticle2018a}, we score the QoIs within all and only the kinetic stages (solid line regions shown in Fig.~\ref{fig: ss_to_t}).  
Similarly, we only score the QoIs contributed by the fluid part within the dashed line regions. 
Intuitively, in kinetic MC, we simulate and score the particle system from $f(x,v,t=0)$ until $f(x,v,t=\infty)$. In the hybrid model, we simulate and score that from $f^j(x,v)$ to ${f^j}'(x,v)$ for the kinetic part, and ${f^j}'(x,v)$ to $f^{j+1}(x,v)$ for the fluid part, with $j=0,1,\ldots,\infty$.

Solving the hybrid model, i.e., Eqs. \eqref{eqn: k part eq} and \eqref{eqn: f part eq} with kinetic MC, 
however, provides neither efficiency gain nor accuracy loss 
compared to solving the original steady-state Eq.~\eqref{eqn: ss-eq short} with kinetic MC implemented in EIRENE~\cite{reiterEIRENEB2EIRENECodes2005a}. They are computationally equivalent. 
In the next section, we adapt KDMC that replaces collisions in the fluid stage (dashed line regions in Fig.~\ref{fig: ss_to_t}) by a diffusive step to this hybrid model, improving computational efficiency while incurring an acceptable loss of accuracy.

\subsection{Hybrid model based on KDMC}\label{sec: hybrid model kdmc}
Making use of KDMC, the collisions in the fluid stage are replaced by a single diffusive step that represents the accumulated effect of many kinetic steps, as explained in Sec.~\ref{sec: dmc}. 
In simulations, it updates the particle state only once, which massively speeds up the simulation in high-collisional regimes. 
When it comes to the QoIs estimation, the trajectory of diffusive steps does not provide information about individual collisions, rendering classical MC estimators inappropriate. 
In other words, solving the fluid part model (Eq.~\eqref{eqn: f part eq}) 
and estimating corresponding QoIs defined in Sec.~\ref{sec: QoIs} using MC estimators with the trajectories provided by diffusive steps, has a large error. 

In this work, we estimate the QoIs contributed by the fluid part using the fluid limit of Eq.~\eqref{eqn: f part eq}, like other hybrid methods~\cite{horstenHybridFluidkineticModel2020a, valentinuzziTwophasesHybridModel2019b, mortierEstimationPostprocessingStep2022}.
Since the diffusive steps in KDMC are governed by the density equation~\eqref{eqn: t-density}, it is reasonable to use the relevant fluid system.
Therefore, we estimate the QoIs by solving the fluid limit of Eq.~\eqref{eqn: f part eq} that is the fluid system derived in Sec.~\ref{sec: fluid system}. Different from other hybrid models, we solve the fluid system by reconstructing its unknown source terms from the kinetic steps.
Particularly, taking the fluid limit of the fluid part equation~\eqref{eqn: f part eq} as in Sec.~\ref{sec: fluid system} yields
\begin{align}
\label{eqn: f density}
    &\partial_x \left(u_pn_f\right)-\partial_x\left(\frac{1}{mR_t}\partial_x \left(T_pn_f\right)\right)=\intv g(\phi_f)dv + \intv s - s_k dv. \\ 
    \label{eqn: f energy}
    &\partial_x\left(\left(\frac{1}{2}mu_f^2+\frac{1}{2}T_f\right)u_fn_f+n_fT_pu_f-\frac{3}{2mR_t}n_fT_f\partial_xT_f\right) = \intv \frac{1}{2}mv^2g(\phi_f)dv + \intv \frac{1}{2}mv^2(s-s_k)dv,
\end{align}
where $n_f$ and $T_f$ are the fluid part neutral density and temperature, respectively. The momentum equation is neglected since it is unnecessary in 1D cases as explained in Sec.~\ref{sec: fluid system m and e}.
The collision operator $g(\phi_f)$ is defined in~\eqref{eqn: coll op} where $\phi_f$ represents the fluid part neutral distribution. 
The source term $s_f$ is replaced by $s-s_k$ for the following reason.
The fluid part of the hybrid model Eq.~\eqref{eqn: f part eq} is uniquely decided by its source term $s_f$, since $\phi_f$ is the unknown variable and other quantities are known.
Thus, its fluid limit is uniquely decided by the moment of $s_f$.
To evaluate the moments of $s_f$, we notice that 
the original source term $s=R_rM_pn_p$ is known, and we can estimate the moments of $s_k$, such that the moments of $s_f$ are estimated according to $s_f=s-s_k$. Therefore, $s_f$ is replaced by $s-s_k$ in~\eqref{eqn: f density}-\eqref{eqn: f energy}. 
To estimate the moments of $s_k$, we rewrite the kinetic part Eq.~\eqref{eqn: k part eq} as
\begin{equation}\label{eqn: s_k}
    s_k = v\partial_x \phi_k - g(\phi_k)= v\partial_x \phi_k - R_{cx}\left(M_p\intv \phi_k dv' - \phi_k\right) + R_i \phi_k.
\end{equation}
Taking the velocity moments of Eq.~\eqref{eqn: s_k} with respect to $[1, \frac{1}{2}mv^2]$, we obtain
\begin{align}
\label{eqn: sk 1}
    &\intv s_k dv = \partial_x m_{1,k} + R_i m_{0, k}, \\
    \label{eqn: sk 3}
    &\intv \frac{1}{2}mv^2 s_k dv = \partial_x \left(\frac{1}{2}m m_{3,k}\right) - \frac{1}{2}R_{cx}\left(T_p^2+mu_p^2\right)n_k + \frac{1}{2}m \left(R_{cx}+R_i\right) m_{2,k},
\end{align}
where the kinetic part moments $m_{l,k}$ with $l=0,1,2,3$ are defined as $\intv v^l \phi_kdv$. They can be easily estimated using MC estimators during the kinetic steps \cite{reiterEIRENEB2EIRENECodes2005a,luxMonteCarloParticle2018a}.
Next, substituting Eqs.~\eqref{eqn: sk 1}–\eqref{eqn: sk 3} into Eqs.~\eqref{eqn: f density}–\eqref{eqn: f energy} closes the system. The QoIs of the fluid parts $n_f, u_f$, and $T_f$ are then obtained by solving the fluid system with $u_f$ solved by Eq.~\eqref{eqn: gamma}. 

One may argue that we can estimate $s_f$ and its moments directly from the definition of $s_f$ in Eq.~\eqref{eqn: f part eq}, namely, $s_f=\sum_{j=0}^\infty\left({f^j}' - f^{j+1}\right)$. 
In fact, it involves estimating quantities at a specific time, for instance at time $t=j\dt$ for the moments of $f^j$. This corresponds to a point estimator, which typically exhibits larger variance than estimators based on quantities averaged over a time interval. For a discussion of this issue, see Chap.~6.IV in \cite{luxMonteCarloParticle2018a} and Appx.~A in \cite{tangAnalysisKineticdiffusionMonte2025}.
More importantly, compared with estimating $s_f$ directly from Eq.~\eqref{eqn: f part eq}, the proposed strategy does not require additional implementation effort. 
Indeed, the moments $m_{l,k}=\int v^l \phi_k\,dv$ with $l=0,1,2,3$, computed for the kinetic part of QoIs, can be reused to estimate $s_f$.
Therefore, the proposed strategy is preferable.


In conclusion, in the proposed hybrid model, we generate particle trajectories using KDMC. 
During the kinetic steps, we estimate the velocity moments $m_{l,k}$ using MC estimators. These moments serve two purposes. First, they provide the moment of the kinetic part, i.e., $m_{l,k}$ with $l=0,1,2$. Second, they yield the moments of the source term of the kinetic part $s_k$, as in Eqs.~\eqref{eqn: sk 1}-\eqref{eqn: sk 3}.
Next, we solve the closed system Eqs.~\eqref{eqn: f density}-\eqref{eqn: f energy} and Eq.~\eqref{eqn: gamma}, with the resulting moments of $s_k$, to obtain the QoIs of the fluid part $n_f, u_f$ and $T_f$. These QoIs give the moments of the fluid part, denoted as $m_{l,f}$, using Eq.~\eqref{eqn: moments}. 
The total moments are therefore
\begin{equation}\label{eqn: total moment}
    m_l=m_{l,k} + m_{l,f}\quad \text{with } l =0, 1, 2.
\end{equation}
Finally, the neutral macroscopic quantities $n,u,$ and $T$ are then calculated as in Eq.~\eqref{eqn: QoIs} with the moments Eq.~\eqref{eqn: total moment}.

\subsection{Comparison with the scheme in~\cite{mortierEstimationPostprocessingStep2022}}
\label{sec: hybrid model comparison}
The scheme in~\cite{mortierEstimationPostprocessingStep2022} considers only charge-exchange collisions and a single density equation. To highlight the differences from the estimation schemes, we adapt this approach to our setting by incorporating ionization processes, and only the single density equation is applied.

Since the diffusive steps of KDMC are governed by the time-dependent density equation~\eqref{eqn: t-density}. Thus, the scheme in \cite{mortierEstimationPostprocessingStep2022} proposed to estimate the QoIs contributed by the diffusive steps via solving Eq.~\eqref{eqn: t-density}. 
Specifically, let $C$ denote a spatial cell of the discretized computational grid.
The starting position of all diffusive steps, ${x^j_i}'$ defined in Eq.~\eqref{eqn: trajectory} with $j=0,\ldots, J_i$ and $J_i$ the total number of time steps of the $i$-th particle,  gives the initial condition, as
\begin{equation}\label{eqn: initial density}
    n_f(x=C,t=0) = \int_C\intv \sum_{j=0}^{\infty} {f^j}'(x,v) dvdx \approx\sum_{i=0}^I\sum_{j=0}^{J_i} w_i\mathbbm{1}_{C}({x_i^j}'),
\end{equation}
where $I$ is the total number of particles, $w_i$ is weight of the $i$-th particle as defined in Sec.~\ref{sec: kmc}, and  $\mathbbm{1}_{C}(x)$ is the indicator function over the cell $C$. 
The simulation time $\hat{\theta}$ is estimated by the average duration of all diffusive steps, that is,
\begin{equation}\label{eqn: averaging time}
    \hat{\theta} = \EX[\theta] \approx \sum_{i=1}^{I}\sum_{j=0}^{J_i} \frac{1}{I} \frac{1}{J_i}\theta_i^j,
\end{equation}
where $\theta$ is the random variable representing the duration of diffusive steps,  
and $\theta_i^j$ is the diffusive time of the $j$-th time step of the $i$-th particle. 
We refer to \cite{tangAnalysisKineticdiffusionMonte2025} for the details. With a homogeneous background, the estimation is unbiased, since the diffusive time $\theta=\dt-\tau$ with $\tau\sim\text{Exp}(R_t)$ for all particles always follows the same distribution. 
Here, $R_t$ is a constant due to the homogeneity. 
However, with a heterogeneous background, like the fusion case, the estimation is biased, as will be shown numerically in Sec.~\ref{sec: exp2 num}.

In our model, we solve the steady-state fluid limit~\eqref{eqn: f density} with the source term~\eqref{eqn: sk 1}, without the time estimation error caused by Eq.\eqref{eqn: averaging time}. 
Unlike the point estimator (more precisely, the analog estimator) used in Eq.~\eqref{eqn: initial density}, the kinetic part moments $m_{l,k}$ with $l=0,1$ in Eq.~\eqref{eqn: sk 1} are estimated using the track-length estimator \cite{luxMonteCarloParticle2018a} along kinetic steps:
\begin{equation}
    m_{l,k} (x=C) \approx \sum_{i=0}^I\sum_{j=0}^{J_i} w_i (v_i^j)^l \int _0^{\tau_i^j} \mathbbm{1}_{C}({x_i^j} + v_i^j t) dt,
\end{equation}
\todo{added the concise formula for the track-length estimator}
where $\tau_i^j$ is the kinetic time in the $j$-th time step of the $i$-th particle. The time integral represents the dwell time that the $i$-th particle inside the cell $C$, given that the velocity $v_i^j$ remains constant during each kinetic step.
With the track-length estimator, a lower variance is expected.
Apart from the model error that both schemes share, our model has an extra discretization error in the source term estimation, since the spatial derivative is present in Eq.~\eqref{eqn: sk 1}.
However, the mesh used for the fluid model is typically sufficiently fine, due to the non-smooth background, so that the discretization error has only a limited effect, see \cite{tangAnalysisKineticdiffusionMonte2025}.

In the numerical experiment presented in Sec.~\ref{sec: exp2 num}, we consider a periodic test case to avoid the effect of the boundary discussed in Sec.~\ref{sec: fluid system n} and Sec.~\ref{sec: bc}, since in \cite{mortierEstimationPostprocessingStep2022}, a large oscillation near the boundary is reported in the neutral velocity defined in \eqref{eqn: QoIs}. The presence of the oscillation needs to be further investigated.
\section{Boundary condition}\label{sec: bc}
\begin{figure}[h]
	\centering
	\begin{tikzpicture}[x=1pt,y=1pt]

        \draw[] (0,0) -- (100, 0);

        \draw (100, -5) -- (100, 5);
        \foreach \y in {-10,-8,-6,-4,-2,0,2,4,6,8} {
            \draw (100,\y) -- (103,\y+1.5);
        }

        \draw (0, -5) -- (0, 5);
        \draw (100, 20) node [anchor=north, inner sep=2pt, color=KULeuvenLichtblauw]  
		{\scriptsize Specular reflective BC};
        \draw (100, -20) node [anchor=south, inner sep=2pt, color=KULeuvenLichtblauw]
		{$x_R$};

        \draw (0, 20) node [anchor=north, inner sep=2pt, color=KULeuvenLichtblauw]
		{\scriptsize Absorbing BC};
        \draw (0, -20) node [anchor=south, inner sep=2pt, color=KULeuvenLichtblauw]
		{$x_L$};
		
	\end{tikzpicture}
	\caption{The one dimension simulation domain $x\in[x_L, x_R]$ with the absorbing BC imposed at $x_L$ and the specular reflective BC at $x_R$.} 
	\label{fig: domain and bcs}
\end{figure}
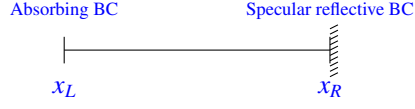

In the plasma edge simulation, as illustrated in Fig.~\ref{fig: domain and bcs}, the upstream boundary is considered to absorb all neutrals, assuming that the neutrals leaving the domain at the upstream position are ionized and never re-enter the domain as neutrals~\cite{horstenComparisonFluidNeutral2016a}.
The downstream boundary at the divertor plate (wall) is reflective. 
In this work, we impose the specular reflective BC at the downstream wall. 
The advanced reflective BC that uses the TRIM database~\cite{horstenDevelopmentAssessment2D2017} and is implemented in EIRENE~\cite{reiterEIRENEB2EIRENECodes2005a} can be readily integrated once the specular reflection condition is clearly specified, see~\cite{horstenComparisonFluidNeutral2016a,horstenDevelopmentAssessment2D2017}. In the following, we discuss the two BCs for the proposed fluid model in Sec.~\ref{sec: fluid bc}, and for KDMC and thus the proposed hybrid model in Sec.~\ref{sec: kdmc bc}.

\subsection{Boundary conditions for fluid system}\label{sec: fluid bc}
Let the simulation domain be defined as $x \in [x_L, x_R]$.
For the fluid system, including Eqs~\eqref{eqn: ss-density}, \eqref{eqn: ss-mom}, and \eqref{eqn: ss-energy}, the specular reflective BC corresponds to enforcing zero normal flux at the boundary. For instance, the particle flux density at $x_R$ (downstream) is
\begin{equation*}
    J_n(x_R) = \left(u_pn - \frac{1}{mR_t}\partial_x T_p n\right)\bigg|_{x=x_R} = 0.
\end{equation*}

For the absorbing boundary condition, the neutral distribution function is known from 
\eqref{eqn: phi appox} and \eqref{eqn: phi appox ns}. 
The corresponding particle, momentum, and energy flux densities at $x_L$ (upstream), therefore, are given by
\begin{equation}\label{eqn: fluid bc absorbing}
    J_n(x_L) = \int_{-\infty}^0 v\,\phi(x_L, v)\,dv, \quad
    J_m(x_L) = \int_{-\infty}^0 m v^2\,\phi(x_L, v)\,dv, \quad
    J_E(x_L) = \int_{-\infty}^0 \frac{1}{2} m v^3\,\phi(x_L, v)\,dv .
\end{equation}
Here, $\phi$ is approximated by~\eqref{eqn: phi appox} for the particle flux, and by~\eqref{eqn: phi appox ns} for the momentum and energy fluxes.

\subsection{Boundary conditions for KDMC}\label{sec: kdmc bc}

\subsubsection{Specular reflective boundary conditions}
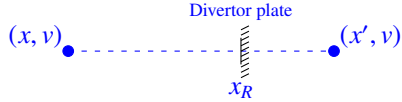
\begin{figure}[hb]
	\centering
	\begin{tikzpicture}[x=1pt,y=1pt]

        \draw[color=KULeuvenLichtblauw,dash pattern={on 2pt off 3pt}] (0,0) -- (100, 0);

        \draw (65, -5) -- (65, 5);
        \foreach \y in {-10,-8,-6,-4,-2,0,2,4,6,8} {
            \draw (65,\y) -- (68,\y+1.5);
        }
        
		\filldraw[color=KULeuvenLichtblauw] (0, 0) circle (2pt);
        \filldraw[color=KULeuvenLichtblauw] (100, 0) circle (2pt);

        \draw (0, 12) node [anchor=north east, inner sep=2pt, color=KULeuvenLichtblauw]
		{$(x,v)$};
        \draw (100, 12) node [anchor=north west, inner sep=2pt, color=KULeuvenLichtblauw]
		{$(x',v)$};
        \draw (65, 20) node [anchor=north, inner sep=2pt, color=KULeuvenLichtblauw]
		{\scriptsize Divertor plate};
        \draw (65, -20) node [anchor=south, inner sep=2pt, color=KULeuvenLichtblauw]
		{$x_R$};
		
	\end{tikzpicture}
	\caption{A particle crosses the divertor plate (physical boundary) at $x_R$ in a diffusive step. $(x,v)$ is the starting state of the diffusive step. The position $x'$ in $(x', v)$, located outside the computational domain, denotes the end position of the particle after the diffusive step. 
    The dashed line indicates that the movement is a diffusive step, where the detailed trajectory is unknown.
    The time that the particle arrives at $x_R$ from $x$ is assumed to be $\tilde\theta = |(x_R-x)|/u_p(x)$. The probability of ionization before hitting the boundary is then $1-\exp(-R_i(x)\tilde\theta)$.} 
	\label{fig: refl d step}
\end{figure}

In kinetic MC, if a particle hits the boundary within a kinetic step, its velocity reverses sign, and it continues to travel with the reflected velocity until the next collision event.
After that collision, a new kinetic step begins.
The kinetic step in KDMC is identical to the kinetic step in standard kinetic MC.
In the diffusive step of KDMC, suppose a particle starts the step from the state $(x,v)$ as shown in Fig.~\ref{fig: refl d step}. Most likely, its end state is $(x',v)$ with $x'$ being out of the simulation domain, when $x$ is close to the boundary.
Physically, however,  the particle might undergo repeated reflections at the boundary, since the ionization rate is low in its vicinity, and when the particle is reflected to re-enter the domain, it is prone to drift towards the boundary again.

To address the out-of-domain issue, \cite{mortierEstimationPostprocessingStep2022} proposed the following strategy.
If the end position of a diffusive step, $x'$, lies outside the simulation domain, the particle is reset to its previous state $(x,v)$. 
Instead of performing the diffusive step, the particle subsequently executes only kinetic steps until ionization occurs.
However, due to the repeated reflections described above, a significant portion of the computational cost is concentrated near the boundary.
Then the speed-up of KDMC compared with kinetic MC is not significant anymore. 
Moreover, significant oscillations are observed in the velocity near the boundary when the out-of-domain issue is treated in this manner.

This inefficiency motivates the development of an alternative strategy. 
Before presenting our scheme, we first introduce an important feature of the hybrid model: the simulation of a particle can be terminated at any time, 
allowing us to balance accuracy and computational efficiency.
Specifically, if we stop the simulation of a particle at, say, time $t$, all kinetic steps of this particle after time $t$ disappear. 
Then all kinetic part quantities, especially the source term $s_k$ and its moments, lose corresponding components, since these quantities are estimated along the kinetic steps. 
In our hybrid model (Eqs.~\eqref{eqn: k part eq} and \eqref{eqn: f part eq}), as explained in Sec.\ref{sec: hybrid model}, the solutions and thus the QoIs are uniquely decided by the source terms $s_k$ and $s_f$.
Hence, the lost kinetic part components are then transferred to the fluid part via the source term $s_f=s-s_k$ and its moments.
By solving the fluid system Eqs.~\eqref{eqn: f density}–\eqref{eqn: f energy} derived from Eq.~\eqref{eqn: f part eq}, the contribution to the QoIs lost from the kinetic part is compensated by the corresponding contribution from the fluid part.
Since kinetic scoring is generally more accurate but more expensive, whereas the fluid reconstruction is cheaper but relies on a closure approximation, this feature can be employed to balance accuracy and efficiency.

Based on this feature, we introduce a reflective-boundary treatment for KDMC with a tunable parameter $\alpha\in[0,1]$. The parameter controls the probability that a particle trajectory is terminated when a diffusive step reaches the reflective boundary, thereby transferring the remaining contribution of that trajectory from the kinetic to the fluid part.

Consider a diffusive step starting from $(x,v)$ whose tentative end position $x'$ lies outside the domain, $x'>x_R$, as illustrated in Fig.~\ref{fig: refl d step}. Since the detailed trajectory during a diffusive step is not resolved, we replace this boundary-crossing diffusive step by a boundary-reaching step consistent with the local diffusive-limit drift: we set $x'=x_R$, sample a new velocity from the local Maxwellian $M_p(\cdot\mid x_R)$, and retain only velocities pointing into the computational domain. The time required to reach the wall is approximated by
\begin{equation}
    \tilde{\theta}=\frac{|x_R-x|}{u_p(x)},
\end{equation}
where $u_p(x)$ is the mean plasma velocity at position $x$, assumed positive close to the boundary. The probability that the particle is ionized before reaching the boundary is then approximated by
\begin{equation}
    1-\exp(-R_i(x)\tilde{\theta}).
\end{equation}

This time approximation is consistent with the diffusive regime near the boundary. In the high-collisional limit, the average time for a particle to travel from $x$ to the wall at $x_R$ is given by~\cite{mortierKineticDiffusionAsymptoticPreservingMonte2022}
\begin{equation}
    \theta = \frac{\mathbb{E}[|x_R-x|]}{u_p(x)}+\mathcal{O}(\varepsilon^2),
\end{equation}
where the $\mathcal{O}(\varepsilon^2)$ term accounts for correlations in the particle-position increments. In the derivation in~\cite{mortierKineticDiffusionAsymptoticPreservingMonte2022}, it assumes that $u_p$ varies slowly over the short distance to the wall and therefore treats it as locally constant.

If the particle survives until it reaches $x_R$, its trajectory is terminated with probability $\alpha$. For $\alpha=1$, particles are terminated immediately upon reaching the reflective boundary, so the boundary contribution is handled almost entirely by the fluid component of the hybrid model; in this case, $s_k(x_R)\approx 0$ and $s_f(x_R)\approx s(x_R)$. For $\alpha=0$, particles are terminated only by ionization, so the reflective boundary is resolved with the maximal kinetic contribution allowed by the KDMC time step $\Delta t$. In the limit $\Delta t\to 0$, KDMC converges to kinetic MC, and this choice recovers the kinetic reflective boundary condition. Intermediate values of $\alpha$ provide a tunable balance between computational efficiency and kinetic resolution near the boundary.

Note that we can combine the idea from \cite{mortierEstimationPostprocessingStep2022} that particles always perform kinetic steps close to the boundary. Then we stop the particle at a certain time. 
This strategy is expected to be more accurate than the one that retains diffusive steps, although it may be computationally more expensive. 
On the other hand, the particle simulation could also be terminated earlier to improve efficiency, but the accuracy is also lost.  
In addition to this, another strategy for the boundary in KDMC is proposed in \cite{steelFluidBoundaryConditions2025}. It modifies the diffusive steps Eq.~\eqref{eqn: SDE update} by incorporating the absorbing and reflective BCs into the solution of the Fokker-Planck equation~\eqref{eqn: op 1}. 
The balance and the comparison among these strategies need to be further investigated. 
In this work, we apply the strategy that retains diffusive steps for exclusively focusing on the hybrid model and the $\alpha$ scheme, and leave such optimization and comparison as future work.

\subsubsection{Absorbing boundary conditions}
With the absorbing boundary condition, in kinetic MC, when a particle arrives at the boundary at $x_L$, it is ionized. The kinetic step in KDMC is again identical to the kinetic step in kinetic MC. For the diffusive step, we treat it again as the kinetic step. Since particles are ionized at the boundary, the strategy that forces particles to stop is unnecessary.

\section{Numerical experiments}\label{sec: num}
In this section, we numerically verify the proposed fluid system and hybrid model using two test cases.
In Sec.~\ref{sec: fusion test case}, we consider a one-dimensional flux tube representative of a detached ITER case~\cite{Kukushkin_2007}. 
We first compare the fluid system, i.e., Eqs~\eqref{eqn: ss-density} and \eqref{eqn: ss-energy} with the single density equation \eqref{eqn: ss-density} and the AFN model \cite{horstenComparisonFluidNeutral2016a} in Sec.~\ref{sec: exp 1 num fluid model}. Then, we evaluate the hybrid model (with the kinetic part Eq.~\eqref{eqn: k part eq} and the fluid part Eqs.~\eqref{eqn: sk 1}-\eqref{eqn: sk 3}) by comparing it with the fluid system, in Sec.~\ref{sec: exp 1 num hybrid}. 
Next, we reduce the charge-exchange collision rate in Sec.~\ref{sec: num1 low col} to verify the performance of the hybrid model when the charge-exchange collision is not dominant.
The absorbing and specular reflective BCs discussed in Sec.~\ref{sec: bc} are considered in this test case.
In Sec.~\ref{sec: exp2 num}, we consider the periodic test case from \cite{maesHilbertExpansionBased2023a} to compare the proposed hybrid model with the QoI estimation scheme for KDMC introduced in \cite{mortierEstimationPostprocessingStep2022} and presented in Sec.~\ref{sec: hybrid model comparison}.

\subsection{1D Fusion test case}\label{sec: fusion test case}

\begin{figure}[h]
    \centering
    \includegraphics[width=0.7\linewidth]{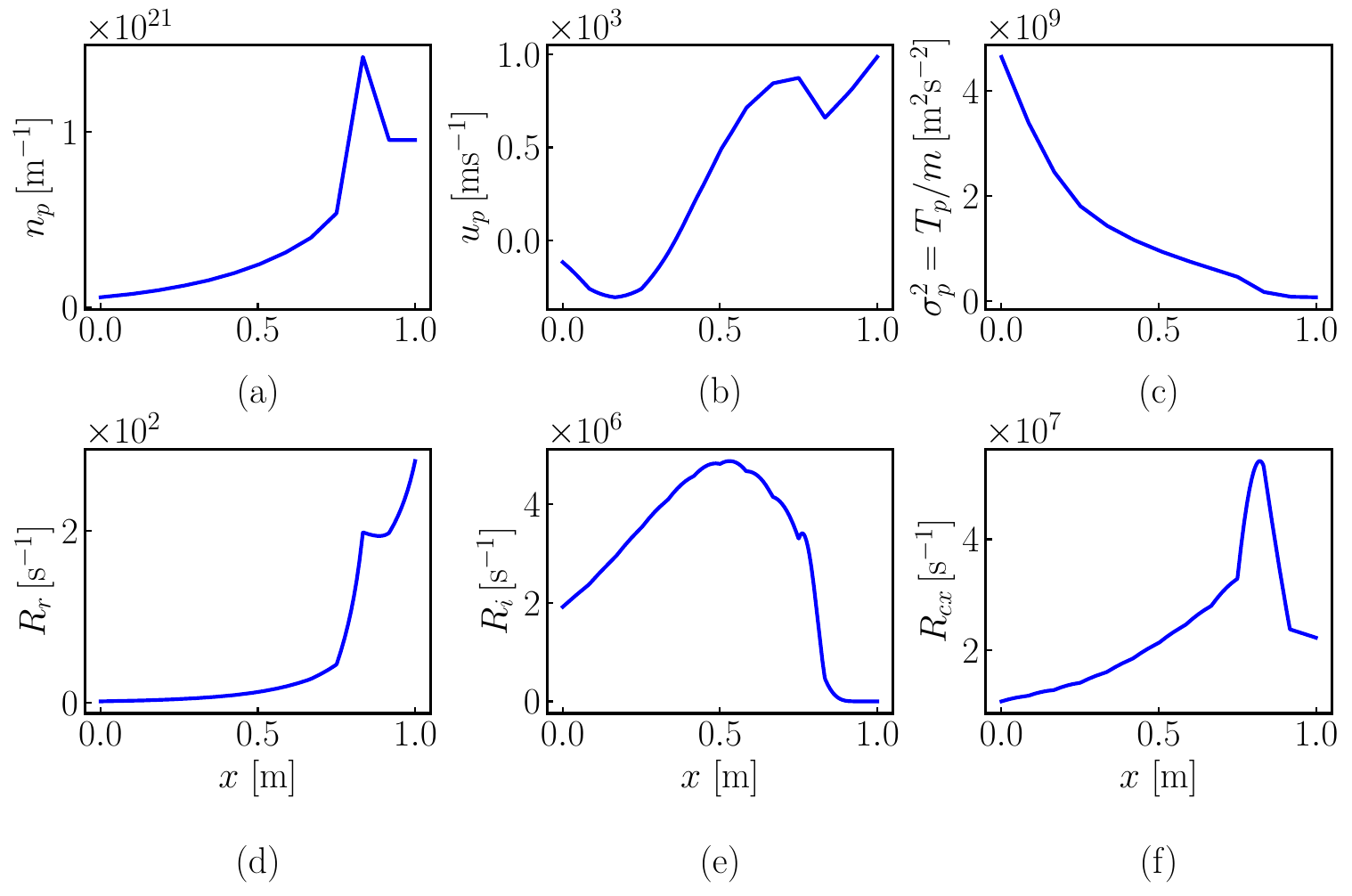}

    \caption{Plasma background used in the simulation. (a) Plasma density $n_p$, where the electron density is assumed equal to the ion density under the quasi-neutrality assumption. (b) Poloidal plasma velocity $u_p$, obtained from the parallel velocity multiplied by the pitch. (c) Velocity variance $\sigma^2_p$, defined as the ion temperature $T_p$ divided by the ion mass. (d) Radiative recombination rate $R_r$. (e) Electron impact ionization rate $R_i$. (f) Charge-exchange rate $R_{cx}$.} 
    \label{fig: background}
\end{figure}
This test case is derived from a two-dimensional SOLPS simulation on the F12 geometry~\cite{Kukushkin_2007} and is reduced to one dimension, with poloidal variation only, see~\cite{horstenComparisonFluidNeutral2016a}. 
The plasma background shown in Fig.~\ref{fig: background} is fixed.   
The simulation domain is $x\in[0,1]$ with the target (divertor) placed at $x=1$.
An absorbing BC is imposed at the upstream boundary $x=0$, while the specular reflection BC is imposed at the downstream boundary $x=1$.
As displayed in Fig.~\ref{fig: background}, close to the downstream boundary, we have $R_{cx}\sim\BO(1/\varepsilon^2)$, $\sigma^2_p\sim\BO(1/\varepsilon^2)$, and $R_i\sim\BO(1)$, which satisfy the diffusive scaling assumption in Sec.~\ref{sec: scaling}. However, the mean velocity of plasma particles is $u_p\sim\BO(1/\varepsilon)$, not $u_p\sim\BO(1)$ as in the assumption. 
As explained in Sec.~\ref{sec: fluid system m and e}, we have modified the distribution approximation accordingly.
Nevertheless, we will see in the following sections that the fluid model still has limitations, which necessitate the use of a hybrid model. 
A similar issue also occurs when the hydrodynamic assumption \cite{horstenComparisonFluidNeutral2016a, maesHilbertExpansionBased2023a} is applied.

The kinetic MC simulation with the track-length estimator is taken as the reference in what follows. 
The simulation uses $I=10^9$ particles; thus, the statistical error is expected to be small, of order $\BO(1/\sqrt{I})=\BO(10^{-4.5})$. 
The simulation domain for the reference solution is uniformly discretized into 400 cells.

\subsubsection{Comparison of fluid models}\label{sec: exp 1 num fluid model}
In this section, we verify the fluid system, including Eqs~\eqref{eqn: ss-density} and \eqref{eqn: ss-energy}. The system is compared with the AFN model and the single-density model whose first and second moments are given in Eq.~\eqref{eqn: moments approx}. In the AFN model, the neutral distribution at the boundary in Eq.\eqref{eqn: fluid bc absorbing} is unknown. As in the original work \cite{horstenDevelopmentAssessment2D2017, horstenComparisonFluidNeutral2016a}, we approximate the neutral distribution via the diffusion theory. We refer to \cite{horstenComparisonFluidNeutral2016a} for the details.

To improve numerical stability, the steady system is solved using the pseudo-time stepping approach, that is, the solution is advanced in artificial time until a steady state is reached. 
Each equation in the system, the single-density model, and the AFN model has the form of
\begin{equation}
    \partial_t H(\psi)+\partial_x(C(\psi)\psi-D\partial_x(\kappa\psi)) = S(\psi),
\end{equation}
where the functions $H, C$ and $S$ depend on the unknown variables $\psi=[n, T]^T$, and $H(\psi)=[n, \frac{1}{2}(mu^2+T)n]^T$ for the density and energy equations, respectively. 
Implicit Euler is used for the time integration. 
The advection term $C(\psi)\psi$ is discretized by a second-order upwind scheme, and the diffusion term $-D\partial_x(\kappa\psi)$ is discretized by the central difference scheme. 
The right-hand side $S(\psi)$ may contain spatial discretization terms. 
For these terms, the MUSCL scheme with the monotonized central (MC) limiter is applied in order to handle the non-smooth and exponentially varying background.
For equations with nonlinear terms, the Newton iteration is employed.

Due to the sharp gradient near the right boundary $x=1$, a very fine mesh is required in this region. 
Based on a uniform mesh with 400 cells for the kinetic MC, we construct a non-uniform fine mesh for fluid models by refining the last 10 cells near $x=1$. Specifically, the $10$-th cell from the boundary is subdivided into $10$ subcells, the $9$-th cell into $11$ subcells, the 8th cell into $12$ subcells, and so on, with the number of subcells increasing toward the boundary.
The fluid model contains both model error (relative to the kinetic model) and discretization error. To ensure that the comparison with the kinetic model is not affected by discretization error, a finer mesh is used to verify that the numerical solution of the fluid model no longer changes significantly.

\begin{figure}[h]
    \centering
    \includegraphics[width=0.9\linewidth]{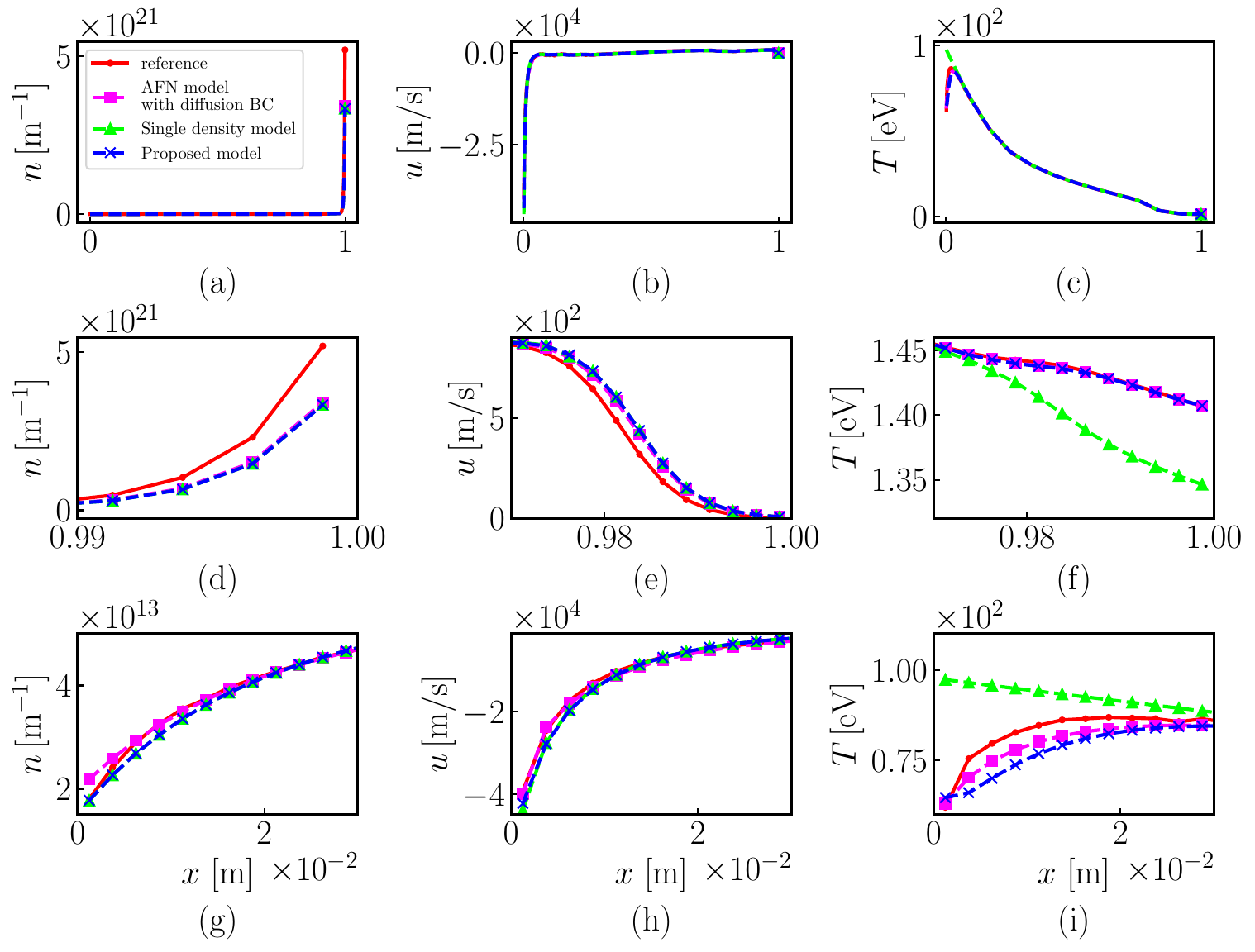}
    \caption{Comparison of fluid models: the proposed fluid system, the single-density model, and the AFN model. 
    Each row shows the neutral density $n$, velocity $u$, and temperature $T$ (from left to right). 
    The first row presents the solutions over the entire domain $x \in [0,1]$. The second and third rows show zoomed views of the solutions near the left and right boundaries, respectively.}
    \label{fig: fluid models comparison}
\end{figure}

The comparison among the three models, the proposed fluid system (blue lines with {\color{blue}$\times$}), the single-density model (green lines with \textcolor{green}{$\blacktriangle$}), and the AFN model with the diffusion BC approximation (magenta line with  {\color[rgb]{1,0,1}$\blacksquare$}), is shown in Fig.~\ref{fig: fluid models comparison}. 
From the first row, we see that all fluid models fit well with the reference simulated using standard kinetic MC (red lines with {\color[rgb]{1,0,0}$\bullet$}) in the interior region, in terms of all QoIs (density, velocity, and momentum). 

Close to the right boundary (second row), where the reflective BC is imposed, the density and velocity from the three models have no obvious difference. 
The density (Fig.~\ref{fig: fluid models comparison}(d)) from the proposed fluid system and the single-density model overlap, since the density equations of these two models are identical. 
The velocity (Fig.~\ref{fig: fluid models comparison}(e)) is computed as 
$u=\Gamma/n$ for all three models, as described at the beginning of this section. Consequently, no significant differences are observed, as the density remains essentially identical across the models.
The temperature (Fig.~\ref{fig: fluid models comparison}(f)) predicted by the single-density model shows a clear deviation from the reference solution. 
As explained in Sec.~\ref{sec: fluid system n}, the second moment from the single-density model in Eq.~\eqref{eqn: moments approx} is derived from the density approximation \eqref{eqn: phi appox} obtained via Hilbert expansion. 
However, this expansion does not capture boundary effects accurately. 
The results, therefore, highlight the necessity of using the proposed fluid model rather than the single-density model.

Near the left boundary (third row), the density (Fig.~\ref{fig: fluid models comparison}(g)) and velocity (Fig.~\ref{fig: fluid models comparison}(h)) predicted by the three models remain largely similar. 
For the temperature, the proposed fluid system and the AFN model with the diffusion BC closely match the kinetic solution, while the single-density model fails to capture the correct trend. 

To compare the three models quantitatively, we compute their relative $L_2$ errors. Rather than evaluating the error over the entire domain, we focus on the regions near the boundaries, since all three models perform well in the interior region, whereas noticeable differences appear near the left and right boundaries.
To this end, we first map the fine meshes used in the fluid model onto the $400$-cell coarse mesh employed in the kinetic Monte Carlo simulations. We then compare the errors in the first 10 cells near the left boundary and the last 10 cells near the right boundary. The result is shown in Tab.~\ref{tab: fluid models comparison}, which follows the result in Fig.~\ref{fig: fluid models comparison}. 

Next, we comment on the computational cost of the proposed fluid system compared with the AFN model. 
The actual cost depends on factors such as the numerical scheme, implementation details, and so on. 
Therefore, a rigorous comparison of computational cost is not provided here. 
Nevertheless, although both models are similar (see \cite{horstenComparisonFluidNeutral2016a}), the proposed fluid system contains more linear terms. As a result, it is reasonable to expect that solving the proposed fluid system is faster than solving the AFN model. 
In our Python implementation, the pseudo-time stepping in this test case requires only about $10$ iterations for the proposed fluid system, compared to several thousands for the AFN model. 

We then conclude that the energy equation in the fluid system \eqref{eqn: ss-energy} is essential for predicting the temperature. 
Moreover, the proposed fluid system using the derived neutral distribution approximation exhibits similar performance to the AFN model, in which the neutral distribution is approximated using diffusion theory.

\begin{table}[h!]
\centering
\caption{Relative $L_2$ errors near the left and right boundaries for the three methods. For the AFN model, the neutral distribution at the boundaries is approximated using diffusion theory.}
\begin{tabular}{lccc ccc}
\hline
 & \multicolumn{3}{c}{Near Left Boundary} & \multicolumn{3}{c}{Near Right Boundary}  \\
\cline{2-4} \cline{5-7}
 & $n$ & $u$ & $T$ & $n$ & $u$ & $T$ \\
\hline
Proposed fluid model & 3.78\% & 9.96\% & 8.15\% & 35.85\% & 19.46\% & 0.12\% \\
Single-density model & 3.78\% & 11.59\% & 18.57\% & 35.75\% & 19.35\% & 3.21\% \\
AFN model & 3.83\% & 4.46\% & 4.67\% & 34.17\% & 16.00\% & 0.10\% \\
\hline
\end{tabular}
\label{tab: fluid models comparison}
\end{table}

\subsubsection{Comparison of hybrid and fluid models}\label{sec: exp 1 num hybrid}
In this section, we verify the proposed hybrid model by comparing it with the proposed fluid system. 
The procedure for solving the hybrid model is concluded in Sec.~\ref{sec: hybrid model kdmc}. 
The scheme with the parameter $\alpha$ for the BC is explained in Sec.~\ref{sec: bc}. Briefly, a larger $\alpha$ indicates a larger portion being simulated by the fluid model at the boundary.

The hybrid method is based on KDMC, and therefore, a time step $\Delta t$ must be chosen.  
According to the analysis for KDMC in~\cite{tangAnalysisKineticdiffusionMonte2025}, the time step should satisfy $\dt\ll\varepsilon^2$ in order to obtain computational gain, where $\varepsilon$ is the diffusive scaling parameter. 
The error is of order $\BO(1/\dt)$ in the regime $\dt\ll\varepsilon^2$.
It is suggested in~\cite{tangAnalysisKineticdiffusionMonte2025} that when the charge-exchange collision rate is of the order of $10^7\,\mathrm{s^{-1}}$, the time step of KDMC should be the order of $10^{-4}$. Here, we choose $\Delta t = 2\times10^{-4}\,\mathrm{s}$, given that the charge-exchange collision rate is around $2\times10^7\sim4\times 10^7$ (see Fig.~\ref{fig: background}).
Besides, $10^6$ particles are used in the hybrid model.

\begin{figure}[h]
    \centering
    \includegraphics[width=0.9\linewidth]{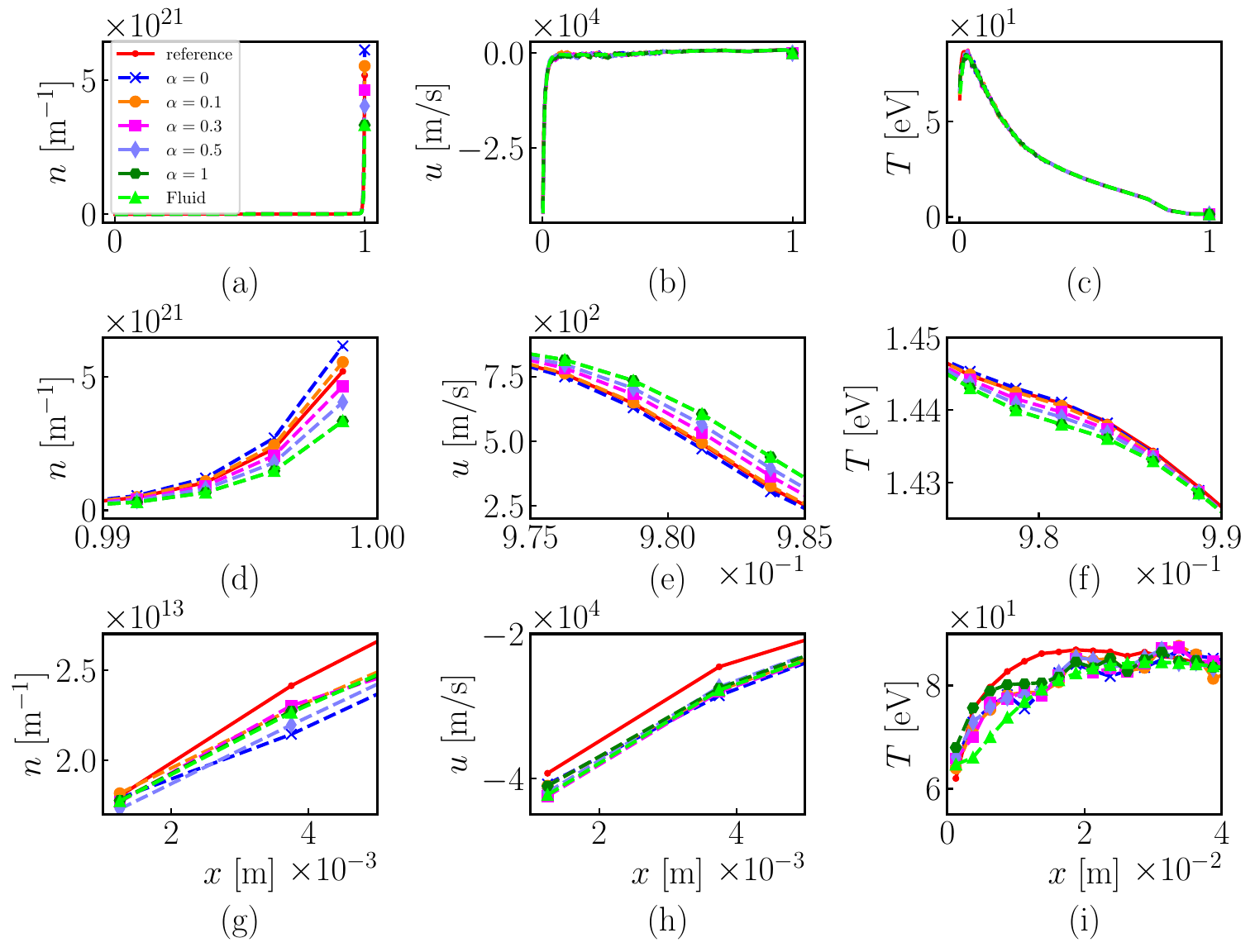}
    \caption{Comparison between proposed hybrid model with $\alpha=0, 0.1, 0.3, 0.5$ and $1$ and the proposed fluid model.
    Each row shows the neutral density $n$, velocity $u$, and temperature $T$ (from left to right). 
    The first row presents the solutions over the entire domain $x \in [0,1]$. The second and third rows show zoomed views of the solutions near the left and right boundaries, respectively.}
    \label{fig: hybrid models comparison}
\end{figure}

The comparison between the hybrid model with $\alpha=0, 0.1, 0.3, 0.5, 1$ and the fluid model is shown in~Fig.~\ref{fig: hybrid models comparison}. 
When $\alpha=1$, particles stop the simulation with a limited number of kinetic steps. 
The right BC is almost handled completely by the fluid part of the hybrid model. 
Thus, the numerical solution of the hybrid model is expected to converge to that of the fluid model. 
This is confirmed in the second row of Fig.~\ref{fig: hybrid models comparison}, where the dark green lines with markers \pyhex{green!50!black} and the light green lines with markers {\color[rgb]{0,1,0}$\blacktriangle$} nearly overlap.
As $\alpha$ decreases, more kinetic steps are retained, and a larger portion of the boundary is treated by the accurate kinetic component of the hybrid model. 
Consequently, the numerical solution approaches the kinetic solution, as shown in the second row of Fig.~\ref{fig: hybrid models comparison} with lines with 
{\color[rgb]{1,0.5,0}\Large$\bullet$},
{\color[rgb]{1,0,1}$\blacksquare$},
{\color[rgb]{0.5,0.5,1}$\blacklozenge$},
and {\color[rgb]{0,0,1}$\times$}.
Note that there are multiple sources of error in the hybrid model, for instance, the KDMC simulation error and the model error, in particular, the error caused by using fluid BCs to approximate the kinetic BCs. 
Therefore, $\alpha=0$ is not always the optimal choice.
The dependence of the optimal $\alpha$ on boundary behaviour and flow regime remains to be investigated.
However, it is already evident that the hybrid model is more accurate than the purely fluid model.
Near the left boundary, as shown in the third row of Fig.~\ref{fig: hybrid models comparison}, the performance of the proposed fluid system and the proposed hybrid model has no obvious difference. 
Besides, in this region, the ionization rate is relatively large (see Fig.~\ref{fig: background} (e)). Particles are ionized quickly, and thus fewer samples are generated. 
Therefore, the stochastic fluctuation in the hybrid model is visible.

The quantitative comparison is shown in Tab.~\ref{tab: hybird model comparison}, where again the solution on the first and last $10$ cells are used to compare under the $L_2$ norm. The comparison confirms the above description for Fig.~\ref{fig: hybrid models comparison}. Especially, when $\alpha=0.3, 0.1$ and $0$, the errors of the hybrid model decrease significantly near the right boundary with the specular reflective BC.
The result demonstrates that the hybrid decomposition systematically recovers kinetic accuracy as $\alpha$ decreases.

Last, in Tab.~\ref{tab: computational cost}, we compare the computational cost from two perspectives.
First, we compare the cost when the same number of particles ($10^7$) is used in both kinetic MC and KDMC.
Within a time step, particles simulated using kinetic MC are expected to collide on the order of $R_t\dt$ times, while particles in KDMC experience at most two collisions. 
Thus, the speedup obtained by using KDMC as the particle simulation scheme is expected to be on the order of $R_t\dt/2$, given that the time of solving the fluid system is negligible.
With $\dt\sim \BO(10^{-4})$ and $R_t\sim\BO(10^7)$, the runtime and speedup are displayed in the first two rows of Tab.~\ref{tab: computational cost}.
Note that, in our treatment of handling the reflective BC, a particle will stop the current time step once it reaches the boundary in a diffusive step. 
Even if the time step has not yet reached $\dt$, a new time step will start. 
Therefore, the computational cost increases as particles have more reflective operations.
In other words, as $\alpha$ decreases, the speedup diminishes. Nevertheless, for all values of $\alpha$, KDMC clearly outperforms kinetic MC in terms of computational efficiency.

Second, we compare the cost for achieving the same level of statistical error. The statistical error for each macroscopic quantity is spatially distributed, and we measure it by taking the relative $L_2$ norm. 
The statistical errors differ for density, velocity, and temperature. We use the velocity error as the reference quantity, since the proposed hybrid model exhibits the smallest reduction in statistical error for velocity compared with the reference kinetic model. The statistical error is computed using Welford’s online algorithm~\cite{welfordNoteMethodCalculating1962} over $100$ trials, with $10^6$ particles in each trial.
The speedup is then displayed in the third row of Tab.~\ref{tab: computational cost}. 
This speedup is larger than that obtained with the same number of particles, since the fluid component of the hybrid model is deterministic. Consequently, the hybrid model achieves a lower statistical error than the pure kinetic model for the same number of particles.

The achieved speed-up of the KDMC approach for this 1D case is much larger than observed for other hybrid methods developed for neutrals in the plasma edge. In \cite{horstenHybridFluidKinetic2018}, it is shown that the speed-up is only a factor of $5$ for the micro-macro method. However, it should be noted that the average simulation time for a single fully kinetic particle trajectory is often artificially large compared to realistic 2D and 3D simulations due to the fact that particles are not able to reach highly ionizing regions in the radial directions. 
The speedup observed in this 1D setting should not be interpreted as a direct estimate for realistic multidimensional simulations, where particles may access highly ionizing radial regions. Nevertheless, the present results indicate that the KDMC-based decomposition may provide substantial computational savings also in higher-dimensional settings.
%
%

\begin{table}[h!]
\centering
\caption{Relative $L_2$ errors near the left and right boundaries for the three methods.}
\begin{tabular}{lccc ccc}
\hline
 & \multicolumn{3}{c}{Near Left Boundary} & \multicolumn{3}{c}{Near Right Boundary}  \\
\cline{2-4} \cline{5-7}
 & $n$ & $u$ & $T$ & $n$ & $u$ & $T$ \\
\hline
Proposed fluid model & 3.78\% & 9.96\% & 8.15\% & 35.85\% & 19.46\% & 0.12\% \\
hybrid model ($\alpha=1$) & 3.07\% & 7.14\% & 4.48\% & 35.88\% & 19.47\% & 0.12\% \\
hybrid model ($\alpha=0.5$) & 3.37\% & 7.85\% & 5.17\% & 22.30\% & 12.44\% & 0.08\% \\
hybrid model ($\alpha=0.3$) & 3.52\% & 10.28\% & 5.92\% & 10.90\% & 7.45\% & 0.05\% \\
hybrid model ($\alpha=0.1$) & 2.83\% & 8.35\% & 5.32\% & 6.47\% & 1.18\% & 0.03\% \\
hybrid model ($\alpha=0$) & 4.62\% & 9.31\% & 6.13\% & 18.01\% & 2.85\% & 0.03\% \\
\hline
\end{tabular}
\label{tab: hybird model comparison}
\end{table}

\begin{table}[h!]
\centering
\caption{Runtime and speedup of kinetic MC and the hybrid model with $\alpha=0, 0.1,0.3,0.5$, and $1$. 
The speedup is defined as the ratio of the runtime of the kinetic MC to that of the hybrid model. The runtime values are truncated to their integer parts.}
\begin{tabular}{lccc ccc}
\hline 
 &Kinetic MC & $\alpha=0$ & $\alpha=0.1$ & $\alpha=0.3$ & $\alpha=0.5$ & $\alpha=1$ \\
\hline
 Runtime with same amount of particles &$1.65\times10^5\,\mathrm{s}$ & $323\,\mathrm{s}$ & $267\,\mathrm{s}$ & $205\,\mathrm{s}$ & $165\,\mathrm{s}$ & $116\,\mathrm{s}$ \\
Speedup with same amount of particles & $-$ & $509.89$ & $616.85$ & $801.25$ & $996.56$ & $1410.50$ \\
Speedup with same statistical error & $-$ & $4286.55$ & $5187.35$ & $2210.34$ & $6282.80$ & $11061.52$ \\
\hline
\end{tabular}
\label{tab: computational cost}
\end{table}

\subsubsection{Hybrid model in non–charge-exchange-dominated regimes}\label{sec: num1 low col}

In this section, we investigate the performance of the proposed hybrid model in regimes where charge-exchange is no longer dominant. The test case is constructed by reducing the charge-exchange rate by a factor of 10, such that the charge-exchange collision rate becomes comparable in magnitude to the ionization collision rate (see Fig. \ref{fig: background}).

\begin{figure}[h]
    \centering
    \includegraphics[width=0.9\linewidth]{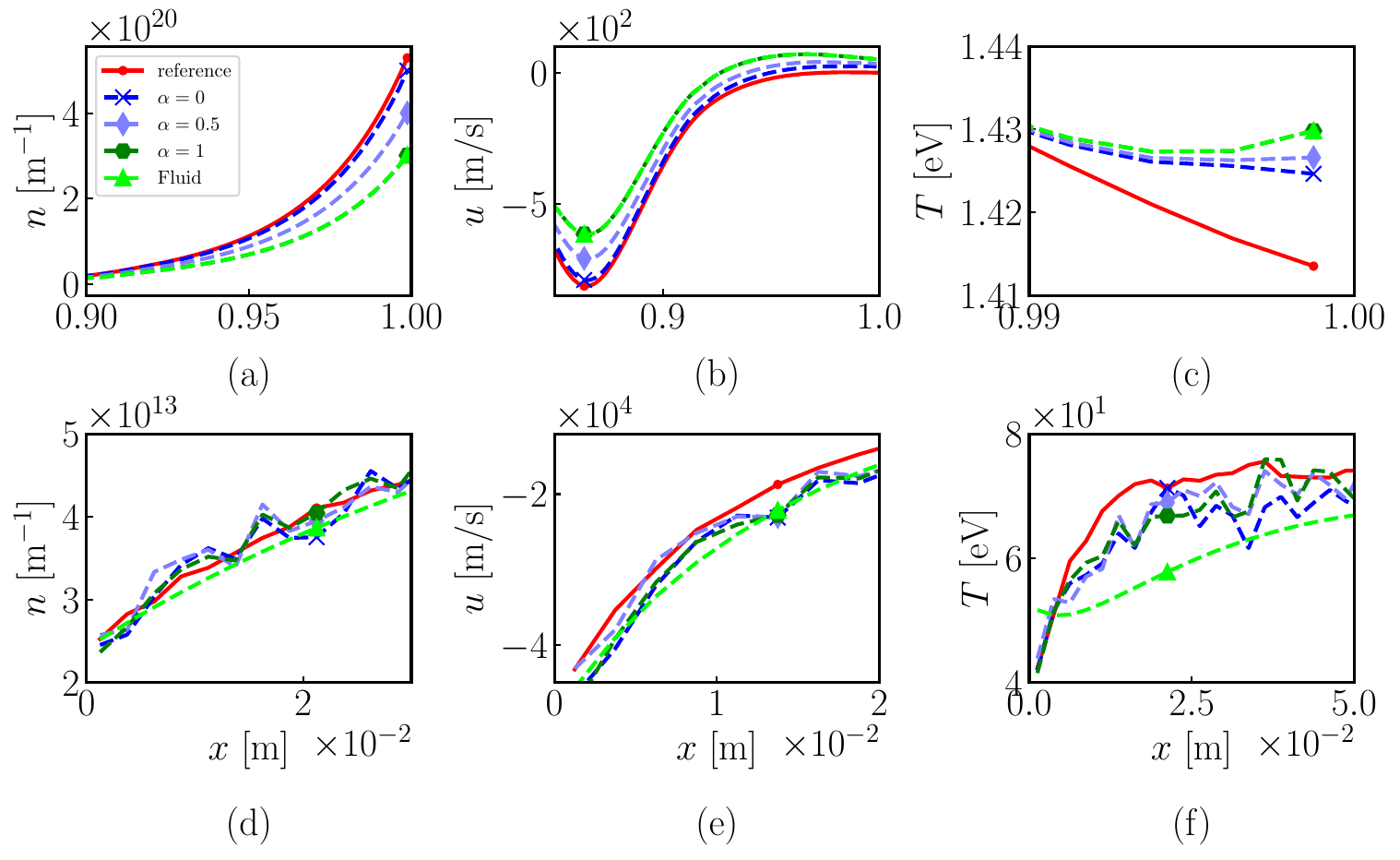}
    \caption{Comparison between proposed hybrid model with $\alpha=0, 0.5$ and $1$ and the proposed fluid model in the non-charge-exchange dominant case.
    Each row shows the neutral density $n$, velocity $u$, and temperature $T$ (from left to right). 
    The solutions over the entire domain are ignored.
    The first and second rows show zoomed views of the solutions near the left ($x=0$) and right ($x=1$) boundaries, respectively.}
    \label{fig: hybrid models comparison low cx}
\end{figure}

When charge-exchange is not dominant, close to the right boundary (the first row in Fig.~\ref{fig: hybrid models comparison low cx}),  the density from the fluid model (the light green line with markers {\color[rgb]{0,1,0}$\blacktriangle$} in Fig.~\ref{fig: hybrid models comparison low cx}(a)) has an obvious difference already starting from $x=0.9$, while in the charge-exchange dominant case, the difference starts from $x=0.99$ (See Fig.~\ref{fig: hybrid models comparison}). 
The significant difference also exists in the velocity in Fig.~\ref{fig: hybrid models comparison low cx}(b).
Nevertheless, the hybrid model can compensate for the discrepancy relative to the reference (the red line with the marker {\color[rgb]{1, 0, 0}$\bullet$}) in both density and velocity. 
Regarding the temperature (Fig.~\ref{fig: hybrid models comparison low cx}(c)), the fluid model exhibits a trend opposite to that of the kinetic reference due to its inherent limitations. 
Although the hybrid model compensates for this discrepancy, a noticeable difference remains even for $\alpha=0$ (light blue line with marker {\color[rgb]{0.5,0.5,1}$\blacklozenge$}) 
where a particle stops the simulation only due to the ionization, 
since the fluid model still contributes through the diffusive steps, 
and therefore affects the hybrid model.
However, this limitation has a reduced impact when more physical BCs are applied.
For instance, in \cite{horstenComparisonFluidNeutral2016a}, an advanced boundary is considered, and the fluid model then has a smaller deviation from the kinetic solution.

Close to the left boundary (the second row in Fig.~\ref{fig: hybrid models comparison low cx}) at which the absorbing BC is imposed, 
the fluid model already has a good performance for density (Fig.~\ref{fig: hybrid models comparison low cx}(d)) and velocity (Fig.~\ref{fig: hybrid models comparison low cx}(e)). For the temperature (Fig.~\ref{fig: hybrid models comparison low cx}(f)), the fluid model still deviates from the kinetic reference, but the hybrid model effectively corrects this discrepancy.

The computational cost is compared in Tab.~\ref{tab: computational cost low-col}. 
In the low charge-exchange collision-rate case, particles have a longer free-flight time, collide less frequently, 
and therefore undergo fewer collisions within a fixed time interval in the kinetic MC method. 
As a result, the KDMC approach yields less computational gain than in the previously considered high-collisional regime. 
Nevertheless, a speedup (with both the same number of particles and the same statistical error as the criteria) of more than 100 times is still achieved.

In conclusion, the proposed hybrid model outperforms the fluid model in this low charge-exchange collision case. 
A discrepancy in the temperature persists near the wall (right boundary). 
This reduced accuracy in this regime is primarily associated with the fluid model near reflective boundaries rather than with KDMC-based decomposition itself. 
It is shown in~\cite{horstenComparisonFluidNeutral2016a} that including more physical BCs will improve the performance of the fluid model. 


\begin{table}[h!]
\centering
\caption{Runtime and speedup of kinetic MC and the hybrid model with $\alpha=0, 0.5$, and $1$. The charge-exchange collisions are not dominant here. The speedup is defined as the ratio of the runtime of the kinetic MC to that of the hybrid model. Runtime values are truncated to their integer parts.}
\begin{tabular}{lccc c}
\hline 
 &Kinetic MC & $\alpha=0$  & $\alpha=0.5$ & $\alpha=1$ \\
\hline
Runtime with same amount of particles  &$1.25\times10^4\,\mathrm{s}$ & $104\,\mathrm{s}$ & $85\,\mathrm{s}$ & $72\,\mathrm{s}$  \\
Speedup with same amount of particles  & $-$ & $120.19$ & $147.05$ & $173.61$  \\
Speedup with same statistical error  & $-$ & $270.42$ & $288.22$ & $390.62$ \\
\hline
\end{tabular}
\label{tab: computational cost low-col}
\end{table}


\subsection{Periodic test case}\label{sec: exp2 num}
In this section, we compare the proposed hybrid model with the scheme in~\cite{mortierEstimationPostprocessingStep2022} for estimating the QoIs defined in Eq.~\eqref{eqn: QoIs}. As discussed in Sec.~\ref{sec: hybrid model comparison}, we use the periodic test case from~\cite{maesHilbertExpansionBased2023a}. 
This test case isolates the estimator-reconstruction component of the hybrid framework from boundary effects. 

\begin{figure}[h!]
    \centering
    \includegraphics[width=0.9\linewidth]{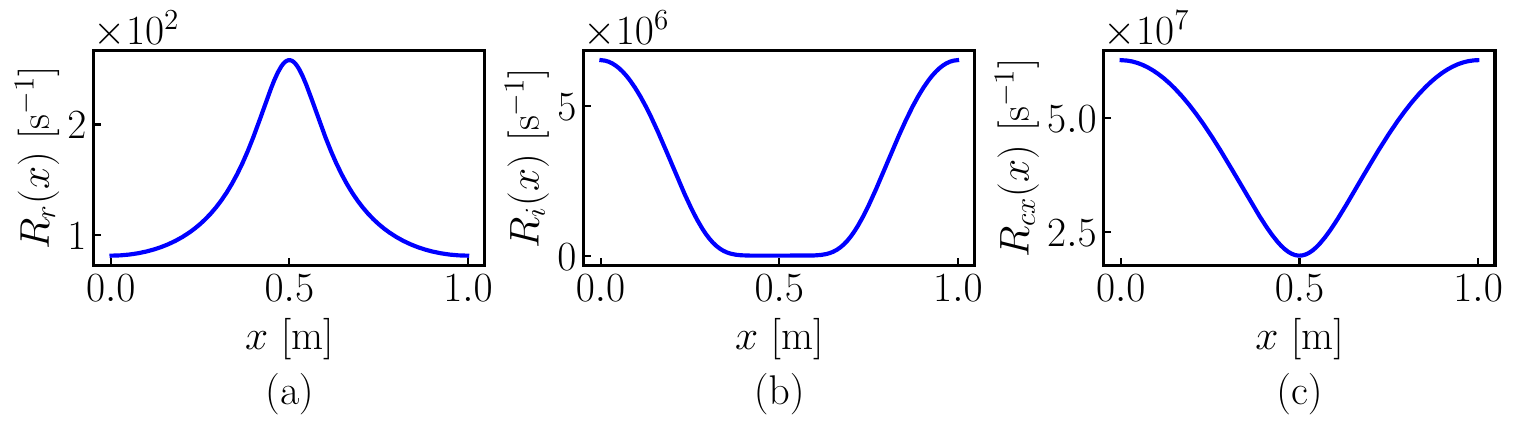}
    \caption{The collision rate profiles of the periodic test case: (a) Recombination rate;  (b) Ionization rate; (c) Charge-exchange rate.
    }
    \label{fig: periodic background}
\end{figure}

In particular, the particle system is simulated on the periodic domain $x\in[0, 1]$, discretized using a uniform mesh with $200$ cells.
The plasma density (or ion density) $n_p=10^{21}\,[\mathrm{m^{-1}}]$ is constant. 
The mean plasma velocity $u_p(x)=0.1\times\sigma_p\,[\mathrm{ms^{-1}}]$ equals $10\%$ of the sound speed, where the variance on the plasma velocity $\sigma_p^2=eT_p(x)/m\,[\mathrm{m^2s^{-2}}]$ with $e\approx1.60\times10^{-19}\,[\mathrm{C}]$ the elementary charge and $m$ the ion mass. 
The temperature $T_p(x)=5.5+4.5\times\cos(2\pi x)$
is taken as a profile that reaches $10\, [\mathrm{eV}]$ at the domain boundaries and equals $1\, [\mathrm{eV}]$ at the center of the domain.
Besides, the recombination $R_r(x)$, ionization $R_i(x)$, and charge-exchange $R_{cx}(x)$ collision rates are displayed in Fig.~\ref{fig: periodic background}. 
Since the background is smooth, no finer mesh is required for the fluid part of the hybrid model. The time step in KDMC is again chosen as $\Delta t = 2\times10^{-4}\,\mathrm{s}$.

Unlike the case with reflective BCs, the system under periodic BCs remains close to equilibrium throughout the entire domain. As a result, in this charge-exchange-dominated case, Hilbert expansion under the diffusive scaling assumption, as described in Sec.~\ref{sec: fluid system n}, remains valid across the whole domain.
Thus, we can adopt the single-density model derived from Hilbert expansion to focus exclusively on differences arising from the estimation procedures. 
Specifically, our scheme solves the steady-state density equation~\eqref{eqn: ss-density}, while the approach in \cite{mortierEstimationPostprocessingStep2022} considers the time-dependent density equation~\eqref{eqn: t-density}. 
In both cases, the first and second moments are estimated using Eq.~\eqref{eqn: moments approx}. Finally, the only distinction between the two approaches lies in the treatment of the density equation, as discussed in Sec.~\ref{sec: hybrid model comparison}.

\begin{figure}[h!]
    \centering
    \includegraphics[width=0.9\linewidth]{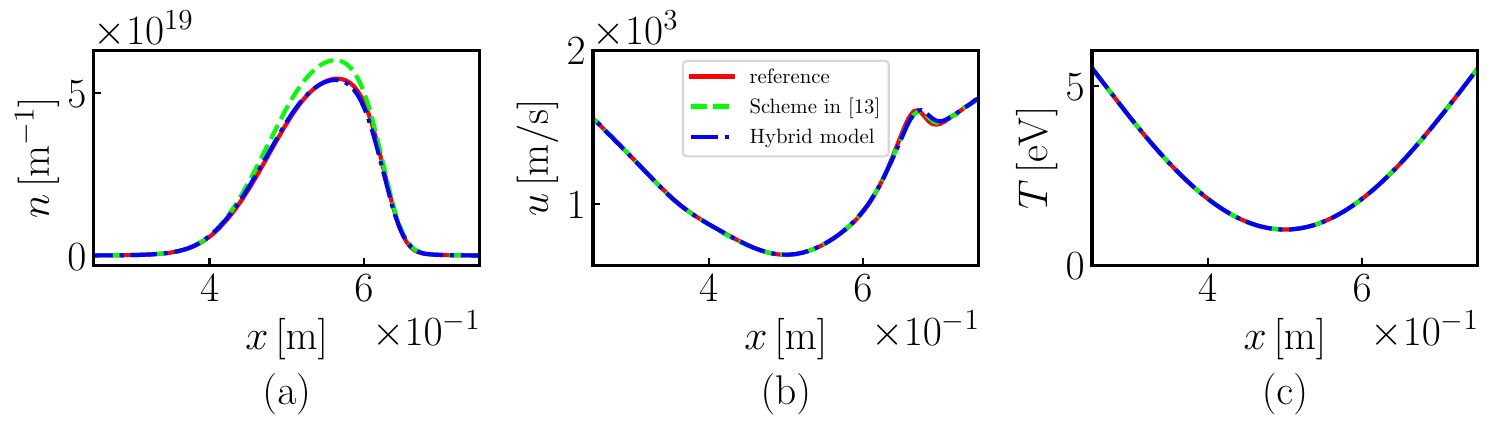}
    \caption{Comparison between the proposed hybrid model and the scheme in~\cite{mortierEstimationPostprocessingStep2022} with a periodic test case.
    (a) the neutral density $n$, (b) velocity $u$, and (c) temperature $T$. 
    }
    \label{fig: hybrid models comparison with Bert}
\end{figure}

The simulation results are shown in Fig.~\ref{fig: hybrid models comparison with Bert}. 
We see that the density (Fig.~\ref{fig: hybrid models comparison with Bert}(a)) computed by the scheme in \cite{mortierEstimationPostprocessingStep2022} (green dashed line {\color[rgb]{0,1,0}\large $--$}) has a larger error than that by the proposed scheme (blue dash-dot line {\color[rgb]{0,0,1}\large $-\cdot$}). 
Since the same numerical scheme is applied to both methods (as explained in Sec.~\ref{sec: exp 1 num fluid model}, the pseudo-time stepping approach is used to solve the steady-state equation) 
and using a finer mesh does not improve the accuracy, 
the error of the scheme in \cite{mortierEstimationPostprocessingStep2022} is then the bias caused by the approximate simulation time $\hat{\theta}$, as explained in Sec.~\ref{sec: hybrid model comparison}. 

\begin{figure}[h]
    \centering
    \includegraphics[width=0.9\linewidth]{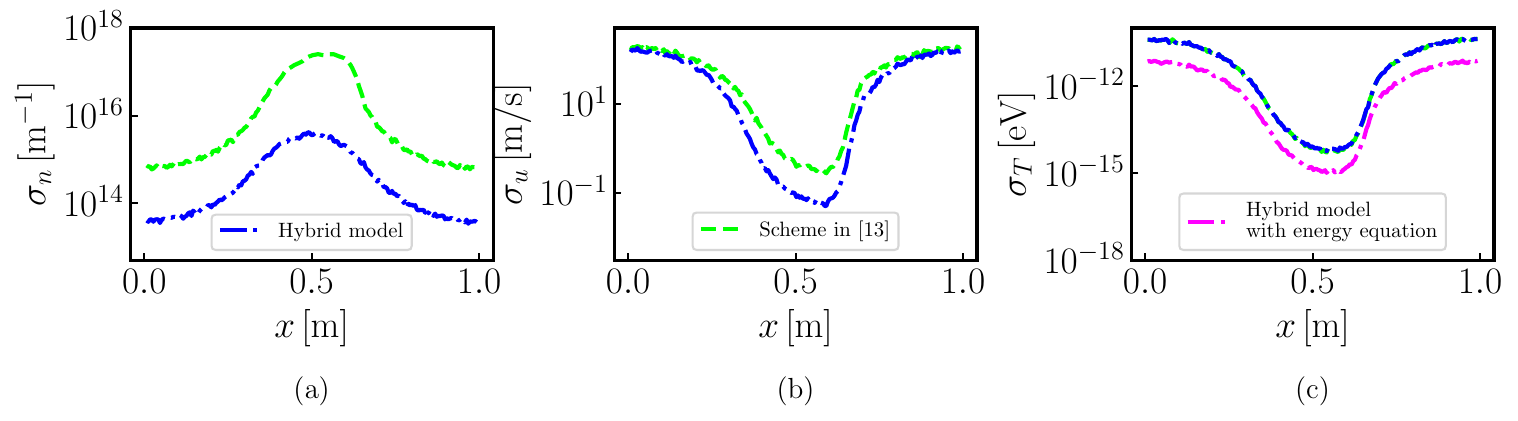}
    \caption{Standard deviations of the proposed hybrid model and the scheme in~\cite{mortierEstimationPostprocessingStep2022} for the periodic test case: (a) standard deviation of density, $\sigma_n$; (b) standard deviation of velocity, $\sigma_u$; and (c) standard deviation of temperature, $\sigma_T$.
    }
    \label{fig: variance}
\end{figure}

Next, we compare the statistical errors of the two schemes, as shown in Fig.~\ref{fig: variance}, where the y-axis is displayed on a logarithmic scale. The errors are computed using Welford’s online algorithm~\cite{welfordNoteMethodCalculating1962} over $100$ trials, with $10^5$ particles used in each trial.
The density obtained with the proposed hybrid model (blue dash-dot line {\color[rgb]{0,0,1}\large $-\cdot$}) exhibits a reduction of at least one order of magnitude in the standard deviation compared with the scheme in~\cite{mortierEstimationPostprocessingStep2022} (green dashed line {\color[rgb]{0,1,0}\large $--$}), as shown in Fig.~\ref{fig: variance}(a).
The difference in the standard deviations of the velocity becomes smaller when the higher-order moment is involved, as illustrated in Fig.~\ref{fig: variance}(b). Furthermore, no significant difference is observed in the temperature standard deviation, as shown in Fig.~\ref{fig: variance}(c).
However, when the energy equation is incorporated into the hybrid model, represented by the magenta dash-dot line {\color[rgb]{1,0,1}\large $-\cdot$}, the standard deviation of the temperature is noticeably reduced.

In conclusion, the proposed model achieves reductions in both bias and variance compared to the scheme in~\cite{mortierEstimationPostprocessingStep2022}. The performance of the hybrid model is further improved when the energy equation is included.
\todo{new plots}

\section{Conclusion}\label{sec: conclusion}
We proposed a distribution-decomposition fluid-kinetic hybrid model based on the asymptotic-preserving kinetic-diffusion Monte Carlo (KDMC) particle simulation scheme. 
The quantities of interest (QoIs) contributed by the kinetic part are estimated using MC estimators, and those by the fluid part are estimated by solving a (steady-state) fluid model, as in other hybrid schemes. 
The source term of the fluid part is reconstructed from the QoIs estimation of the kinetic part. Thus, no extra cost is required.
Compared with the previous strategy that estimates the quantities of interest for KDMC in~\cite{mortierEstimationPostprocessingStep2022}, the proposed hybrid model has smaller bias and variance. The numerical result is also more stable.
Compared with other hybrid models used in plasma edge simulations, the kinetic and fluid parts of the proposed model are not coupled, and thus no iteration scheme is required. 
Consequently, the speedup of using the KDMC-based hybrid model is significant.

To improve the accuracy of the fluid model, based on Hilbert-Chapman-Enskog expansions, we propose a fluid system tailored for KDMC, including density and energy equations in 1D cases.  
It has more linear terms compared with the advanced fluid neutral (AFN) model used in the SOLPS-ITER code suite. The numerical results show that the proposed fluid model has a comparable error to the AFN model, but the computational cost is lower. 

Absorbing and specular reflective boundary conditions are considered. 
For the latter, we introduce an $\alpha$-scheme to further accelerate the simulation. 
The method exploits a key property of the hybrid model, i.e., particle trajectories can be terminated at arbitrary times. With the tunable scheme, one can balance computational efficiency and accuracy.

Numerically, for the charge-exchange-dominated fusion test case, the hybrid model achieves a speedup of at least a factor of 500 compared to the reference kinetic MC method, while maintaining relative $L_2$ errors around $10\%$ near the boundary for the density, velocity, and temperature.
When charge exchange is not dominant, the accuracy becomes more sensitive to the reflective-boundary treatment, primarily due to limitations of the fluid model near the boundary.
It is expected that when more physical BCs are imposed, the fluid model performs more accurately.
Nevertheless, the hybrid model compensates largely for the discrepancy of the fluid model from the reference solution, and still achieves a speedup exceeding a factor of 100 when the collision rate is relatively low.

As future work, improved treatments of reflective boundary conditions should be developed for low-collisional regimes. In particular, it may be necessary to mitigate the influence of the fluid model near the boundary.
\todo{This could be a master's thesis or a PET paper}
The KDMC-based decomposition framework is expected to extend to higher-dimensional settings, provided the corresponding fluid closures and boundary treatments are derived and assessed.
Lastly, the numerical comparison with other hybrid approaches remains an open problem.



\section*{Acknowledgements}
This work has been carried out within the framework of the EUROfusion Consortium, funded by the European Union via the Euratom Research and Training Programme (Grant Agreement No 101052200 — EUROfusion). The views and opinions expressed herein do not necessarily reflect those of the European Commission.
Part of this research was funded by the Research Foundation Flanders (FWO) under grant G085922N. 
The authors are also grateful to Thijs Steel, Vince Maes, and Klaas Willems for valuable discussions.

\section*{CRediT authorship contribution statement}
\textbf{Zhirui Tang:} Conceptualization, Methodology, Software, Investigation, Visualization, Writing – Original Draft.  
\textbf{Niels Horsten:} Supervision, Methodology, Validation, Writing – Review \& Editing.  
\textbf{Giovanni Samaey:} Conceptualization, Supervision, Funding Acquisition, Writing – Review \& Editing.

\section*{Data availability}
The data and source code that support the findings of this study are publicly available at:  
\url{https://gitlab.kuleuven.be/numa/public/kdmc-based_hybrid_model.git}

\bibliographystyle{elsarticle-num-names} 
\bibliography{cas-refs}

\end{document}